\documentclass[manuscript]{aastex}

\usepackage{natbib}

\shorttitle{Chemical modeling of Infrared Dark Clouds}
\shortauthors{Vasyunina et al.}

\begin{document}


\title{Chemical modeling of Infrared Dark Clouds: the Role of
	Surface Chemistry}


\author{T. Vasyunina, A. I. Vasyunin, Eric Herbst\altaffilmark{1}}
\affil{Department of Chemistry, University of Virginia,
              Charlottesville, VA 22904 USA}

\and

\author{H. Linz}
\affil{Max Planck Institute for Astronomy (MPIA), K\"onigstuhl 17, D-69117 Heidelberg, Germany}
%\email{linz@mpia.de}
\altaffiltext{1}{Also: Departments of Astronomy and Physics,
University of Virginia, Charlottesville, VA 22904 USA}


%\altaffiltext{1}{Visiting Astronomer, Cerro Tololo Inter-American Observatory.
%CTIO is operated by AURA, Inc.\ under contract to the National Science
%\altaffiltext{2}{Society of Fellows, Harvard University.}
%\altaffiltext{3}{present address: Center for Astrophysics,
%    60 Garden Street, Cambridge, MA 02138}
%\altaffiltext{4}{Visiting Programmer, Space Telescope Science Institute}
%\altaffiltext{5}{Patron, Alonso's Bar and Grill}


\begin{abstract}
We simulate the chemistry of infrared dark clouds (IRDCs) with a model in which the physical
conditions are homogeneous and time-independent.  The chemistry is solved as a function
of time with three networks: one purely gas-phase, one that includes accretion and
desorption, and one, the complete gas-grain network,  that includes surface chemistry
 in addition.   We compare our results with observed molecular abundances for two
  representative IRDCs --   IRDC013.90-1 and IRDC321.73-1  -- using the molecular species
N$_2$H$^+$, HC$_3$N, HNC, HCO$^+$, HCN, C$_2$H, NH$_3$ and CS.   IRDC013.90-1 is a
cold IRDC, with a temperature below 20 K, while IRDC321.73-1 is somewhat warmer, in the range 20 - 30 K.
We find that the complete gas-grain model fits the data very well, but that the
goodness-of-fit is not sharply peaked at a particular temperature.
Surface processes are important for the explanation of the high gas-phase abundance of N$_2$H$^+$  in IRDC321.73-1.  
The general success of the 0-D model in reproducing single-dish observations of our limited sample of 8 species
shows that it is probably sufficient for an explanation of this type of data.
To build and justify more complicated models, including spatial temperature and density
structure, contraction, and heating, we require high-resolution interferometric data.

\end{abstract}





\section{Introduction}

Based on line and continuum observations as well as supporting simulations, 
the evolutionary stages of low-mass stellar and planetary formation are relatively
well understood.  The molecular inventory and chemistry of these stages -- cold  starless cores,
pre-stellar cores, Class-0 sources and hot corinos, outflows, and protoplanetary disks -- have been extensively investigated
both observationally and theoretically
\citep[e.g.,][]{1997ApJ...486..316B,2004A&A...422..159W,
2007ARA&A..45..339B,2008ApJ...674..984A,2009ApJ...691.1459V, 2009ARA&A..47..427H}.  
%There are many different chemical models including complicated
%physical processes such as contraction and warming up at levels through hydrodynamics in complexity.
%There is much observational data to justify these complex models.
The large amount of high-quality observational data makes the 
construction of complex spatial models of low-mass protostars feasible with chemistry, 
dynamics, and radiative transfer often combined.

In the field of massive star-formation, the overall situation is not as well developed, 
partially because massive young stellar objects exist in complex regions and tend to 
be more distant.  The best studied stage is doubtless the hot core, which is a warm 
region (200 K - 300 K) in which the chemistry differs strongly from the cold envelope.  
This chemistry has been studied for at least two decades, with the current consensus 
that the complex organic molecules detected in the gas are formed mainly on and in 
grain ices both before and during the warm-up stage 
\citep{1988MNRAS.231..409B,1993ApJ...408..548C,2001ApJ...546..324R,2004MNRAS.354.1141V,2008ApJ...682..283G}.   
Nor have only complex  molecules been studied; \citet{2011A&A...525A.141B} 
have investigated the NH$_{3}$/N$_{2}$H$^{+}$ abundance ratio towards the 
high-mass star-forming region AFGL 5142 and shown that the abundance ratio 
is strongly affected by the high temperatures.  The physical structure of hot 
cores has been studied by \citet{2004A&A...414..409N}, while some physical models 
have been developed for starless cores that might pertain to the high-mass 
limit \citep{2008ApJ...683..238K,2010MNRAS.402.1625K}.  
Recent work on the chemistry of intermediate-mass Class 
0 protostars has also been reported concerning CO depletion 
and N$_{2}$H$^{+}$ deuteration \citep{2010A&A...518A..52A}.

 % \citet{2012arXiv1201.4430S} studied the DNC/HNC abundance ratio in 18 high-mass sources, including infra-red dark clouds (IRDCs) and high-mass protostellar objects (HMPOs), and found that the DNC/HNC ratio in some IRDCs is lower than that in HMPOs.  Their chemical simulation showed that the ratio depends to some extent on the physical conditions and history of the sources.
%The main reason for the paucity of such chemical studies has been the absence of multi-molecular observational data to
%test the models.   In this paper, we will use recent observational results on a variety of molecules as the basis for a series of chemical models of IRDCs. These models are simple pseudo-time-dependent 0-D ones, in which the density and temperature remain fixed and homogeneous as the chemistry progresses.}

Unlike high-mass protostars, infra-red dark clouds (IRDCs) are thought to represent the earliest stages of massive stars.  These distant objects were discovered 15 years ago with the
\emph{Infrared Space Observatory \/} \citep[\emph{ISO\/};][]{1996A&A...315L.165P} and
\emph{Midcourse Space Experiment\/} \citep[\emph{MSX\/};][]{1998ApJ...494L.199E}
through their appearance as dark silhouettes against the bright Galactic mid-infrared background.
Detected IRDCs show a broad range of masses from several dozens to several thousands of solar masses
\citep{2000ApJ...543L.157C,2006ApJ...641..389R,2009A&A...499..149V},
densities from $10^{5}$ cm$^{-3}$  to $10^{6}$ cm$^{-3}$ \citep{1998ApJ...508..721C,2009ApJ...705..123G},
and typical temperatures from 8 K to 20 K 
\citep{1998ApJ...508..721C,2006A&A...450..569P,2011ApJ...733...44D,2011ApJ...736..163R}.
Since their initial discovery, several molecular line
observations have been published for IRDCs
\citep[e.g.][]{1998ApJ...508..721C,2006ApJS..166..567R,2010ApJ...721..222B,2009ApJ...702.1615C,2011ApJ...739L..16B,2012arXiv1201.4430S}. 
This allowed the construction of the first models  
\citep[e.g.,][]{2009ApJ...705..123G,2012arXiv1201.4430S}. 
Still, the results of these models could only be compared with a few observed
molecules, and the models contained a simpler physical structure than their low-mass
starless analogs.
Only recently have molecular surveys, in which many lines were observed
for the same object, started to appear in the literature
\citep{2008ApJ...678.1049S,2010ApJ...714.1658S,2011A&A...527A..88V,2011arXiv1108.4446F}.
During the Mopra molecular line survey, \citet{2011A&A...527A..88V}
searched for 37 points within southern IRDCs  using rotational lines of 13 molecular species.
These data can be analyzed to give molecular columns and abundances, 
and these quantities can be
compared with the results of chemical simulations, which is a goal of our research.  

Because IRDCs are cold and dense, at least as seen through the beam of 
single-dish telescopes, they might be expected to resemble low-mass 
starless cores in their chemistry.  Yet their temperature range and 
overall density both tend to be higher, and there is much less knowledge 
of the physical structure of these distant sources.   So, it is useful to 
study their chemistry as distinct objects and then compare salient aspects 
of it with both low-mass prestellar cores, generally of somewhat lower 
temperature and density, and high-mass protostellar objects, generally of 
higher temperature and density.

Although most IRDCs studied to date have temperatures  below 20 K,
we can distinguish some IRDCs with higher temperatures.
For the present study,  we selected two clouds for chemical study: IRDC013.90-1 from the colder group and IRDC321.73-1 from the warmer group.
Both clouds have a simple
morphology at mid-IR, sub-millimeter, and millimeter wavelengths and have been studied with the most molecular species.
Furthermore, the kinematic distance of both regions is clearly below 3 kpc and thus at the lower end of distances for the Mopra sample.

The paper is organized in the following manner. In Section~2 we introduce the IRDCs to be modeled,
while in Section~3, we discuss the chemical model itself.
In Section~4, observational and modeled molecular abundances are compared.
We discuss the chemistry in some detail in Section~5 and present our conclusions in Section~6.

\section{Observational data}

%% In a manner similar to \objectname authors can provide links to dataset
%% hosted at participating data centers via the \dataset{} command.  The
%% second curly bracket argument is printed in the text while the first
%% parentheses argument serves as the valid data set identifier.  Large
%% lists of data set are best provided in a table (see Table 3 for an example).
%% Valid data set identifiers should be obtained from the data center that
%% is currently hosting the data.
%%
%% Note that AASTeX interprets everything between the curly braces in the
%% macro as regular text, so any special characters, e.g. "#" or "_," must be
%% preceded by a backslash. Otherwise, you will get a LaTeX error when you
%% compile your manuscript.  Special characters do not
%% need to be escaped in the optional, square-bracket argument.

\begin{figure}
\centering
	\includegraphics[width=10cm]{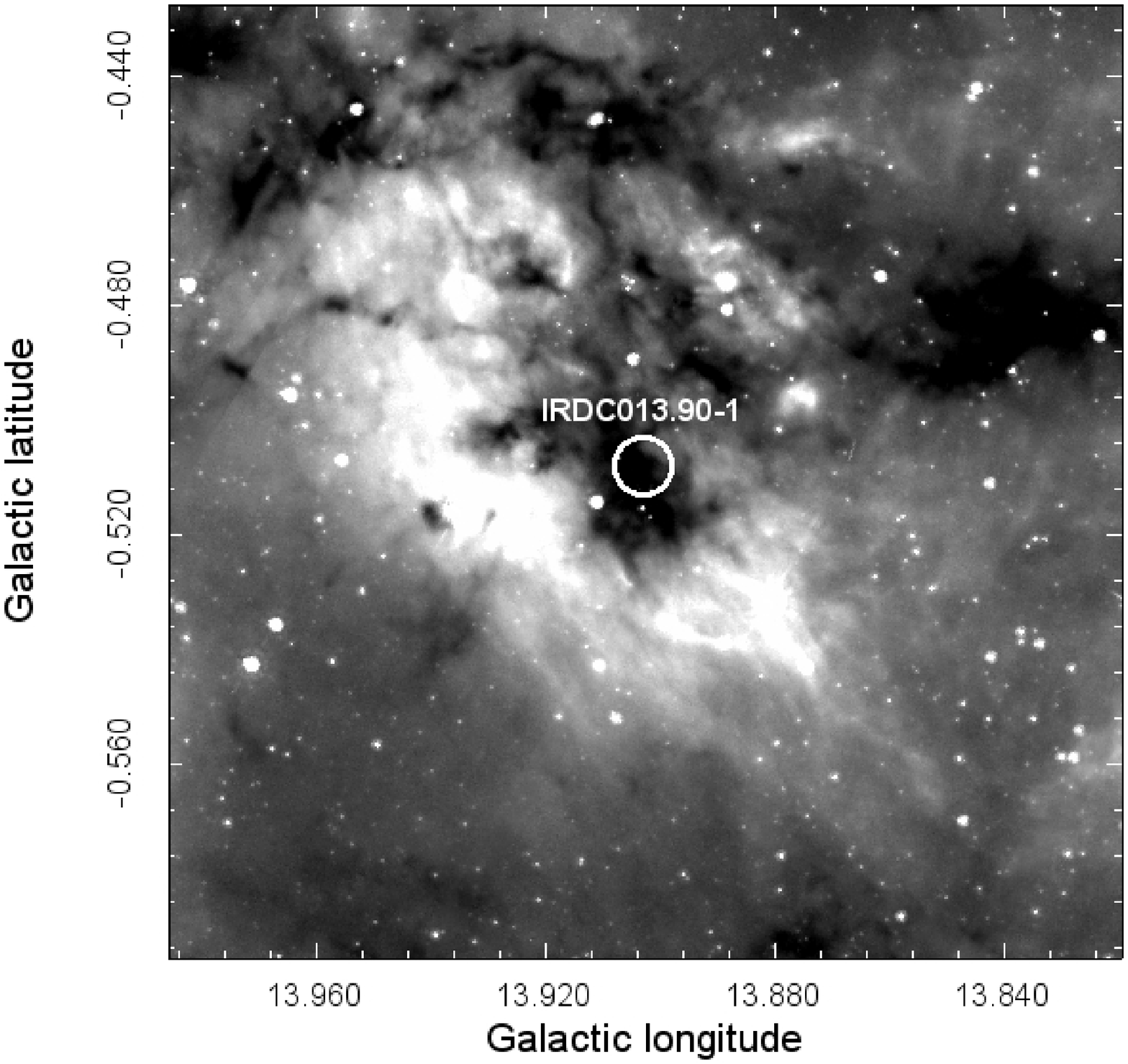}
	\includegraphics[width=10cm]{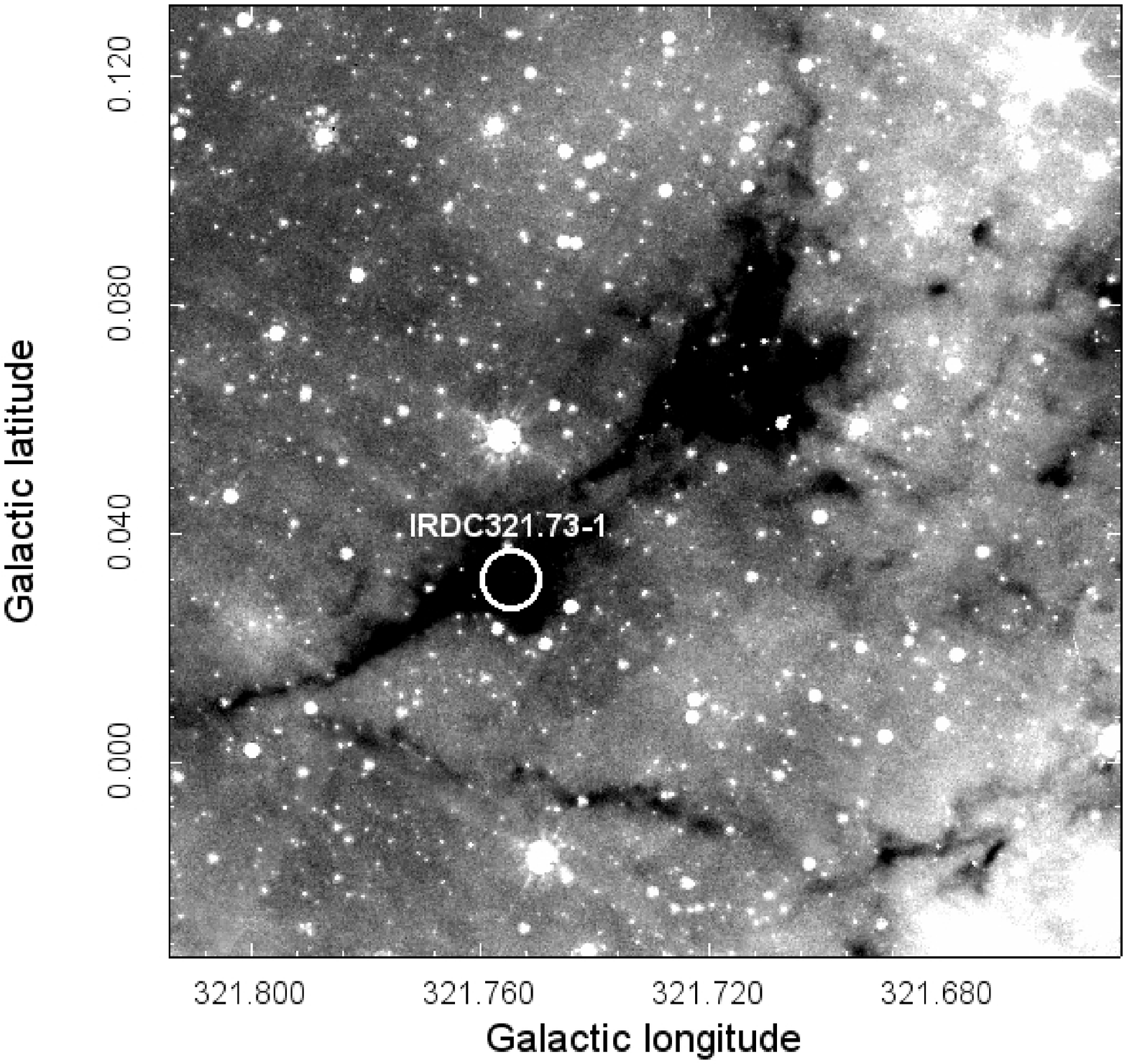}
	\caption{Upper panel: 8 $\mu$m Spitzer/Glimpse image of the IRDC013.90-1. Lower panel:  8 $\mu$m Spitzer/Glimpse image of the IRDC321.73-1.
		Circles mark Mopra observed positions by \citet{2011A&A...527A..88V} and show the beam size.}
\label{figure:3color}
\end{figure}

For our modeling study, we chose
IRDC013.90~-~1 ($\alpha = 18^h17^m33^s$, $\delta=-17^{\circ}06'36''$) and
IRDC321.73~-~1 ($\alpha = 15^h18^m26^s$, $\delta=-57^{\circ}22'00''$),
the observation of which was reported in the Mopra survey by \citet{2011A&A...527A..88V}.
IRDC013.90~-~1 is a part of a large complex (Figure~\ref{figure:3color}, upper panel)  and
IRDC321.73~-~1 is an eastern component of IRDC 321.73+0.05, as shown in  Figure~\ref{figure:3color}, lower panel.
Both clouds show compact rather than filament structure at millimeter, submillimeter and mid-infrared wavelengths.
As part of the program to investigate dust properties in southern IRDCs,  \citet{2011A&A...527A..88V} estimated their masses and
H$_2$ column densities based on 1.2 mm continuum data. For IRDC013.90~-~1 the mass is ~$\approx  430 M_{\sun}$, while for IRDC321.73-1
the mass is ~$\approx  110 M_{\sun}$; the masses are more than enough
to form not only low- and intermediate--mass stars, but also one or two early B stars. 
In comparison with other clouds from the list,  IRDC013.90~-~1 and
IRDC321.73-1 have a quite high H$_{2}$ column density  of $3.5 \times 10^{22}$ cm$^{-2}$ and $3.2 \times 10^{22}$ cm$^{-2}$
respectively \citep{2009A&A...499..149V}.
Such a high column density, considering  that it is derived from single-dish observations with a large beam,
can lead to the absence of any emission at 3.6 - 8 $\mu$m; indeed, only at 24 $\mu$m
can very weak sources be detected  in IRDC013.90~-~1 and IRDC321.73-1.
Given an angular cloud size of about 10$''$, we can transform column densities into
volume densities, which are $\approx 10^5$ cm$^{-3}$.

The main distinction between the two chosen clouds is their ammonia kinetic temperature. In the case of
IRDC013.90~-~1 it is 13 K, while in the case of IRDC321.73~-~1 it is 22 K, which is higher than ``normal''.
From the astrophysical point of view, a 9 Kelvin difference may not be dramatic; however, this difference can
be important for chemical processes on grain surfaces, the rates of which are exponentially sensitive to temperature.

For the present study we used molecular abundances for
N$_2$H$^+$, HC$_3$N, HNC, HCO$^+$, HCN and C$_2$H from the Mopra molecular line survey 
\citep{2011A&A...527A..88V}.  These abundances were obtained 
using the H$_2$ column density estimated from 1.2 mm SIMBA/SEST data from \citet{2009A&A...499..149V}, 
adopting the corresponding beam size.
To add to the molecular data for modeling purposes, we included NH$_3$ data and CS molecular abundances  from other observations.
Ammonia observations were performed with the Parkes telescope (Linz et al. in prep.), 
where both IRDC013.90~-~1 and IRDC321.73-1 were covered.
CS(2-1) observations for IRDCs from the fourth Galactic quadrant were performed by \citet{2008ApJ...680..349J} and do not include IRDC013.90~-~1
from the first quadrant.
To calculate the CS column density for IRDC321.73-1 from the available integral intensity,
we used Eq. 1 from \citet{2011A&A...527A..88V}.
Furthermore, we estimated the CS and NH$_3$ molecular abundances  
in the same way as for other species -  adopting the H$_2$ column density from  
1.2 mm SIMBA/SEST data \citep{2009A&A...499..149V}, 
with the corresponding beam size. 
Observed  abundances with respect to H$_2$ for all species used 
in this study are listed in Table~\ref{table:abundances}, along with calculated abundances described later in the text.

%% In this section, we use  the \subsection command to set off
%% a subsection.  \footnote is used to insert a footnote to the text.


%% This section also includes several of the displayed math environments
%% mentioned in the Author Guide.

\section{Model}

For the chemical simulations,  we utilized a pseudo--time--dependent 
0-D model,  in which the density and temperature remain fixed and 
homogeneous as the chemistry progresses  from initial abundances to a 
final steady-state condition, where for all species in the model, the processes 
of formation and destruction balance out. 
Thus, we consider a stationary model in the sense that it does not include any  large-scale
dynamical 
processes such as warm-up or collapse,  although we do consider adsorption 
and desorption between gas and grains.  The term ``pseudo-time-dependent'' 
does not mean that the chemistry arrives at a steady-state within a reasonable time; 
indeed, for gas-grain models, steady-state conditions are not easily achievable, 
 and can take as long as 10$^{8-12}$ yr to reach, depending upon the 
efficiency of non-thermal desorption, at which time much of the molecular 
material involving heavy elements can lie in grain mantles \citep{2007A&A...467.1103G}.

As we show below, to justify more complicated models, more advanced observational data are
required. In the present study, we use molecules that all trace approximately the same
temperature and density conditions. Another observational limitation is spatial resolution. 
Single-dish observations of the clouds located at several kpc distance do not allow 
the resolution of structures that are smaller than 0.5 pc. Thus any dynamical processes that happen 
on a smaller scale will be smoothed over the beam size. Under such circumstances, the use 
of a 1-D model, where temperature and density are different at different radii from the core 
center, is premature.

In this study, rate equations were employed
together with  three different networks of reactions based on 
the KIDA  
 network release described in
\citet{2010A&A...522A..42S}, which are in turn based on the Ohio State University (OSU) gas-grain chemical network.
The first  network is one in which all surface processes and coupling between gas and dust
are removed except for the surface process of formation of H$_2$ via the reaction
H + H $\rightarrow$ H$_2$ and ion-grain collisions.
The second network adds accretion onto and desorption from grains to the first modification.
 We include two types of desorption  in these networks:  thermal evaporation, which begins to be important at 
 temperatures $\geq$ 15K, when desorption of the most vaporizable species affects 
 the gas-phase chemistry, and cosmic ray desorption, which can dominate
at temperatures  $<$ 15K \citep[see Figure 8 in][]{2007ARA&A..45..339B} . 
Photodesorption is not included in the model for two reasons.  First, the 
 objects considered in this paper are sufficiently opaque,  with A$_{\rm v} >10$, 
 so that the rate of photodesorption via external photons is small  \citep{2008ApJ...672..629V}.  Secondly,
cosmic ray-induced photons cause significant photodesorption only at times greater than
 10$^6$ yr.   Nor do we include desorption via chemical reactions, such as the heat generated by H$_{2}$ formation.The third network used is the full network, which includes all gas-phase and grain surface reactions.  Although
desorption via exothermic reactions on the surface can occur when this network is used in a model  \citep{2007A&A...467.1103G},
its effect is strong mainly for complex molecules, which are not considered here.
Further in the text we refer to the three networks and the models based on them as ``gas-phase", 
``accretion/desorption", and ``surface" networks, respectively.

Starting with  ``low-metal''  initial abundances 
\citep{1974ApJ...193L..35M, 1982ApJS...48..321G, 2008ApJ...680..371W}, shown in Table~\ref{table:initial},
we ran the code with several sets of gas densities and temperatures.
All other parameters, listed in Table~\ref{table:initial_cond}, were fixed.
We assume a cosmic ray ionization rate for H$_{2}$ of 1.3$\times  10^{-17}$~s$^{-1}$,
a granular radius of $10^{-5}$ cm, a granular mass density of 3.0 g cm$^{-3}$, a
visual extinction of 10, and  a gas-to-dust mass ratio of 100,
all of which are typical values for dark clouds.  

In this study, we considered only
diffusive (Langmuir-Hinshelwood)
surface chemical reactions involving physisorbed species.  The monolayers of ice that form are treated as a
fully reactive phase, which leads to what is known as a ``two-phase'' model \citep{1992ApJS...82..167H}.
We assume that the diffusion of species on the surface is  caused only by thermal hopping, and that quantum
tunneling is inefficient even for the lightest species \citep{1999ApJ...522..305K}.
The diffusion/desorption energy ratio ($E_{\rm b}/E_{\rm D}$) is chosen to be 0.5, which is
somewhere in between the extreme estimates of 0.3 \citep{1992ApJS...82..167H} and 0.77 \citep{2001MNRAS.322..770R}.
The assumed site density is
1.5$\times$ 10$^{15}$ cm$^{-2}$, typical for an olivine surface \citep{1999ApJ...522..305K}.

To compare the agreement among model results and  observations, we used two criteria.
First, we simply counted the number of species $C(t)$ for which calculated abundances are in agreement
with observation to within an order of magnitude at time $t$.  Since there are 8 observed species for IRDC321.73~-~1, 
and 7 for IRDC013.90~-~1, the maximum values of $C(t)$ for each cloud are 8 and 7, respectively.
Assuming that all lines except those of  HCO$^+$ are optically thin, and that the ammonia kinetic
temperature equals the excitation temperature, we achieve a factor of a few for the
uncertainties in the observational values \citep{2011A&A...527A..88V}.
But, taking into account uncertainties in the rate coefficients and initial
abundances in the model, we can easily reach a one order-of-magnitude total error range
\citep{2004AstL...30..566V,2008ApJ...672..629V,2005A&A...444..883W,2006A&A...451..551W,2010A&A...517A..21W}.
Secondly, for every species $i$ we estimate a confidence parameter $k{\rm _i}(t)$ from \citet{2007A&A...467.1103G}:

\begin{equation}
k_i(t) = erfc \left( \frac{|log(\chi_{obs})-log(\chi_{mod})|}{ \sqrt{2} \sigma} \right) ,
\label{Equ:confidence_garrod}
\end{equation}
where $\chi_{obs}$ is the observed abundance and
$\chi_{mod}$ is the calculated abundance, $\sigma=1$, and
$erfc$ is the complementary error function for species $i$ ($erfc = 1 - erf$) ; with this definition $k_i(t)$ ranges between zero and unity.  A
calculated value that lies 1 order of magnitude from $\chi_{obs}$ has
a confidence parameter of 0.317, while a value 2 orders of magnitude
from $\chi_{obs}$ has a confidence parameter of 0.046.
%We assume the same one order of magnitude higher and lower limits and hence $\sigma$ equals 1.
Unlike the choice of \citet{2007A&A...467.1103G}, as global confidence parameter $K(t)$ for the whole set of species
we use the lowest $k_i(t)$ value instead of the average. Thus, we estimate the confidence parameter for the species that shows the
worst agreement between model and observations,  and hence, all other molecules exhibit better agreement between model and observations. In this study, we have a limited set of observed molecules to be compared with model, so that it is important to reproduce abundances well for all molecules.   Our choice for a global confidence parameter allows it to be high only in the case where all species are reproduced well.

\section{Results}
\label{section:results}

We have run models with all three networks over a range of temperatures for two
H$_{2}$ densities -- 10$^{5}$ cm$^{-3}$ and 10$^{6}$ cm$^{-3}$.
The temperatures used range from 10 K to 40 K, with 5 K steps between 10-30 K.
In this section,  we emphasize  results for  conditions relating to
``cold'' and ``warm'' IRDCs. 

\subsection{A ``Cold" IRDC:  IRDC013.90~-~1}
\label{section:cold}

\begin{figure}
	\centering
	\includegraphics[width=5cm,angle=90]{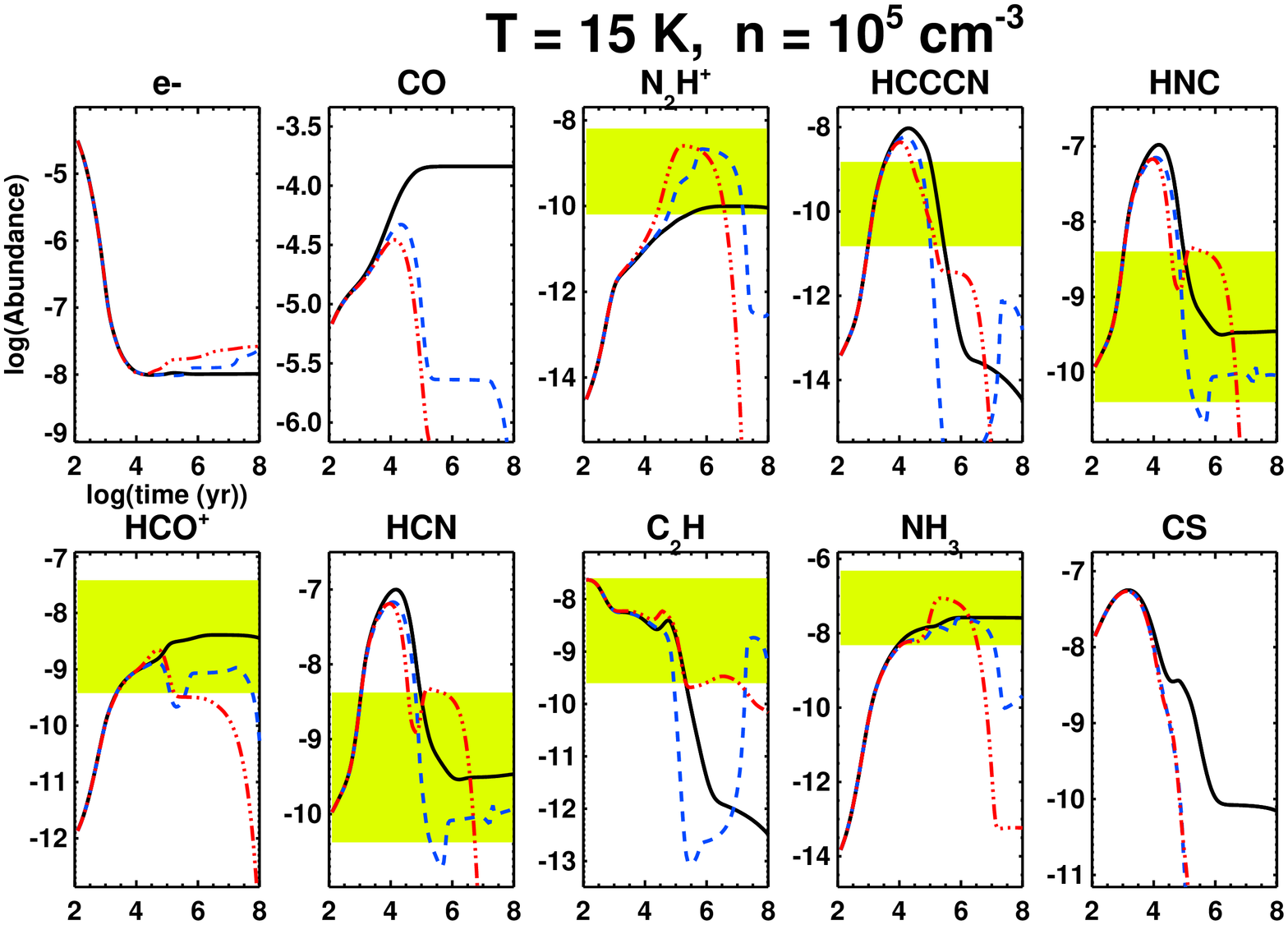}
	
	\includegraphics[width=5cm,angle=90]{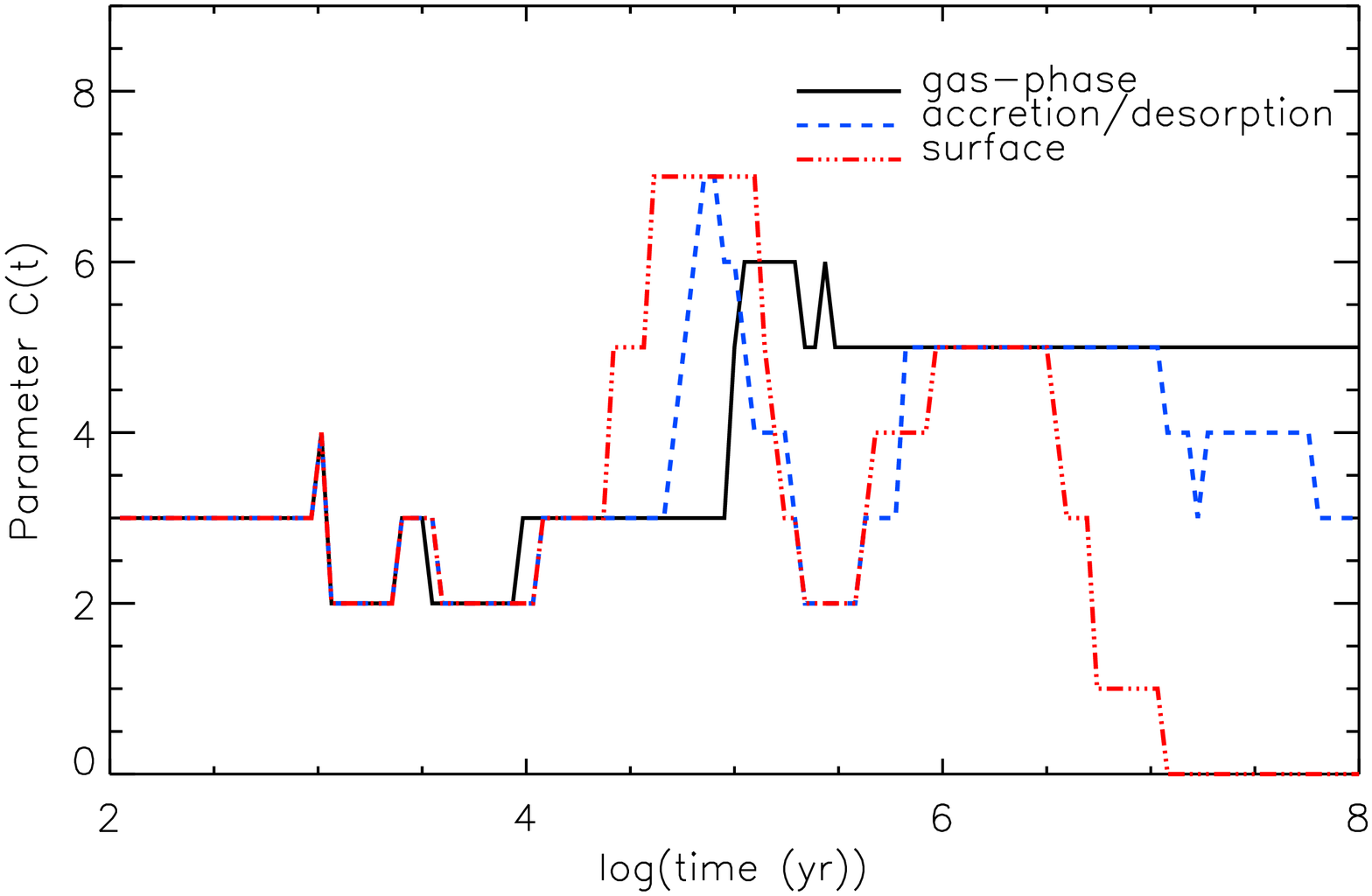}
		
	\includegraphics[width=5cm,angle=90]{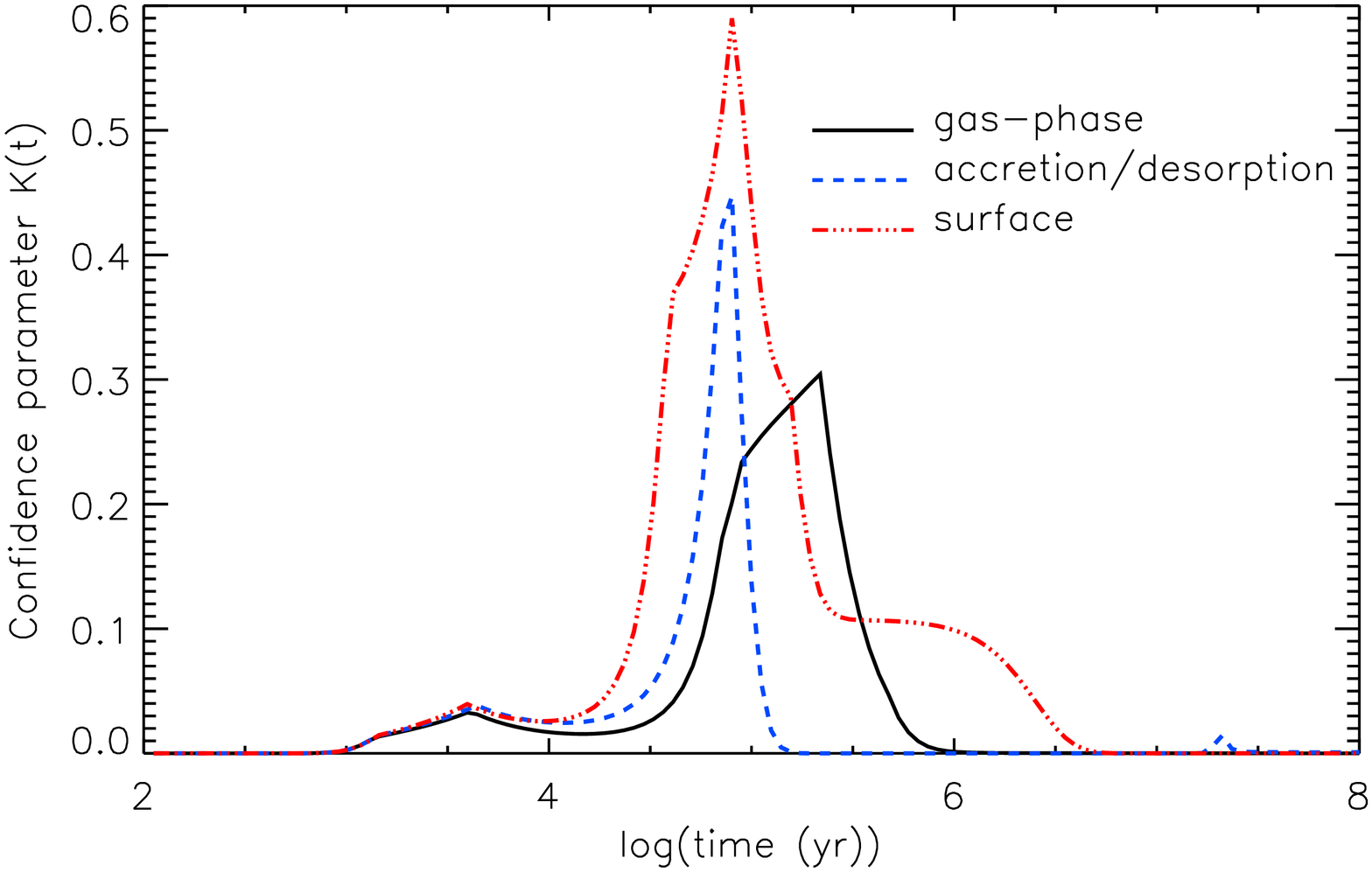}
	
	\caption{Topmost panels: abundance profiles relative to H$_{2}$ as a
		function of time. 
		The  box on every panel corresponds to the observational
		values for IRDC013.90 $\pm$ one order of magnitude.
		Middle panel:  time dependence of the parameter $C(t)$ for the three different networks.
		Lowest panel: time dependence of the confidence parameter $K(t)$
		for the three different networks. 
		 On the topmost panels, the solid black line corresponds to the gas-phase network, dashed  line to 
		the accretion/desorption network, and dash-dotted line to the surface network.
		(A color version of this figure is available in the online journal.)}
\label{figure:15K_106}
\end{figure}

Figure~\ref{figure:15K_106} (topmost 10 panels) shows modeled abundance profiles at times through 10$^{8}$~yr
 at $T = 15$~K and $n(H_{2}) = 10^5$~cm$^{-3}$  for ten species including CO, CS and electrons, although CO has not been studied in the two sources,
 and CS data are not available for IRDC013.90~-~1. 
 The boxes in the panels represent observational values for IRDC13.90-1 with respect to H$_2$ $\pm$ one order of magnitude.
With both the  accretion/desorption and surface networks,  CO freeze-out, which occurs after 10$^{3.5}$ yr,   plays an important
role,  as will be discussed in the next section.  Indeed, the complete freeze-out of CO is prevented mainly by cosmic ray
desorption. As gaseous CO is lowered in abundance, the abundance of  HCO$^+$ is diminished, while
the abundance of N$_2$H$^+$ increases and for a time rises
to more than an order of magnitude higher than the abundance
for the gas-phase network before diminishing.

The value of $C(t)$ for the 7 observed species is shown in the middle portion of  Figure~\ref{figure:15K_106}.  Both the  surface and
accretion/desorption networks  fit all 7 available observational abundances simultaneously although the time range of agreement
for the surface network, $6-9 \times 10^4$ yr, is  much longer.
With the gas-phase network,  a maximum $C(t)$ of 6 occurs at a somewhat later time,
as both N$_{2}$H$^{+}$ and C$_2$H take turns being only slightly outside the observational limits.
The highest (best) value of the
confidence parameter $K(t)$  occurs  for  the surface network, in the vicinity of  $10^5$ yr,
as can be seen in the bottom panel of Figure~\ref{figure:15K_106}.
So, at the temperature and density used, the complete gas-grain network does a marginally better job than the other two.  Modeled abundance values are presented in Table~\ref{table:abundances} for the surface network with $n = 10^{5}$ cm$^{-3}$ and $T = 15$~K at a time of $8 \times 10^{4}$~yr, when the confidence
parameter $K(t)$ reaches its maximum.

Were the physical conditions for IRDC013.90~-~1 not measured, would our simulations enable us to determine them unambiguously?  Even if we only consider the surface network, the situation is murky.  
 In Table~\ref{table:Ct}, we list the maximum $C(t)$ and $K(t)$ values for the three networks at temperatures
 from 10 K to 30 K at 5 K intervals and H$_{2}$ densities of 10$^{5}$~cm$^{-3}$ and 10$^{6}$~cm$^{-3}$.   One can see that a $C(t)$ parameter of 7 is achieved with the surface network for all temperatures at the
 lower density and for two temperatures at the higher density.  If we use the parameter $K(t)$, we see
 that equal fits are obtained for the lower of the two densities at 15 K (the observed
 temperature) and 25 K.   We can conclude only that the lower density is slightly
 preferred, in agreement with observations.

\subsection{A ``Warm" IRDC:  IRDC321.73~-~1}
\label{section:best_fit}

\begin{figure}
	\centering
	\includegraphics[width=5cm,angle=90]{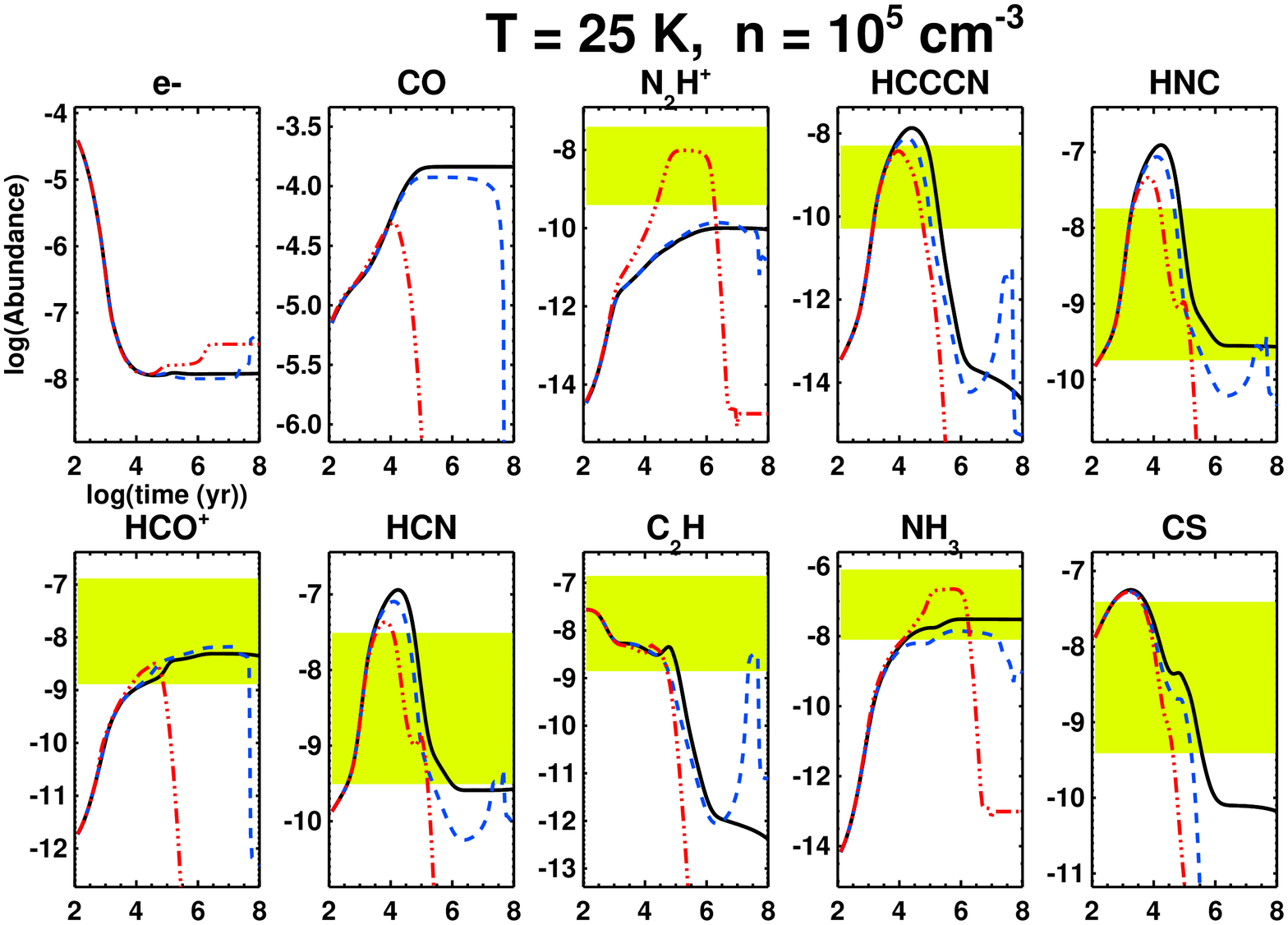}
		
	\includegraphics[width=5cm,angle=90]{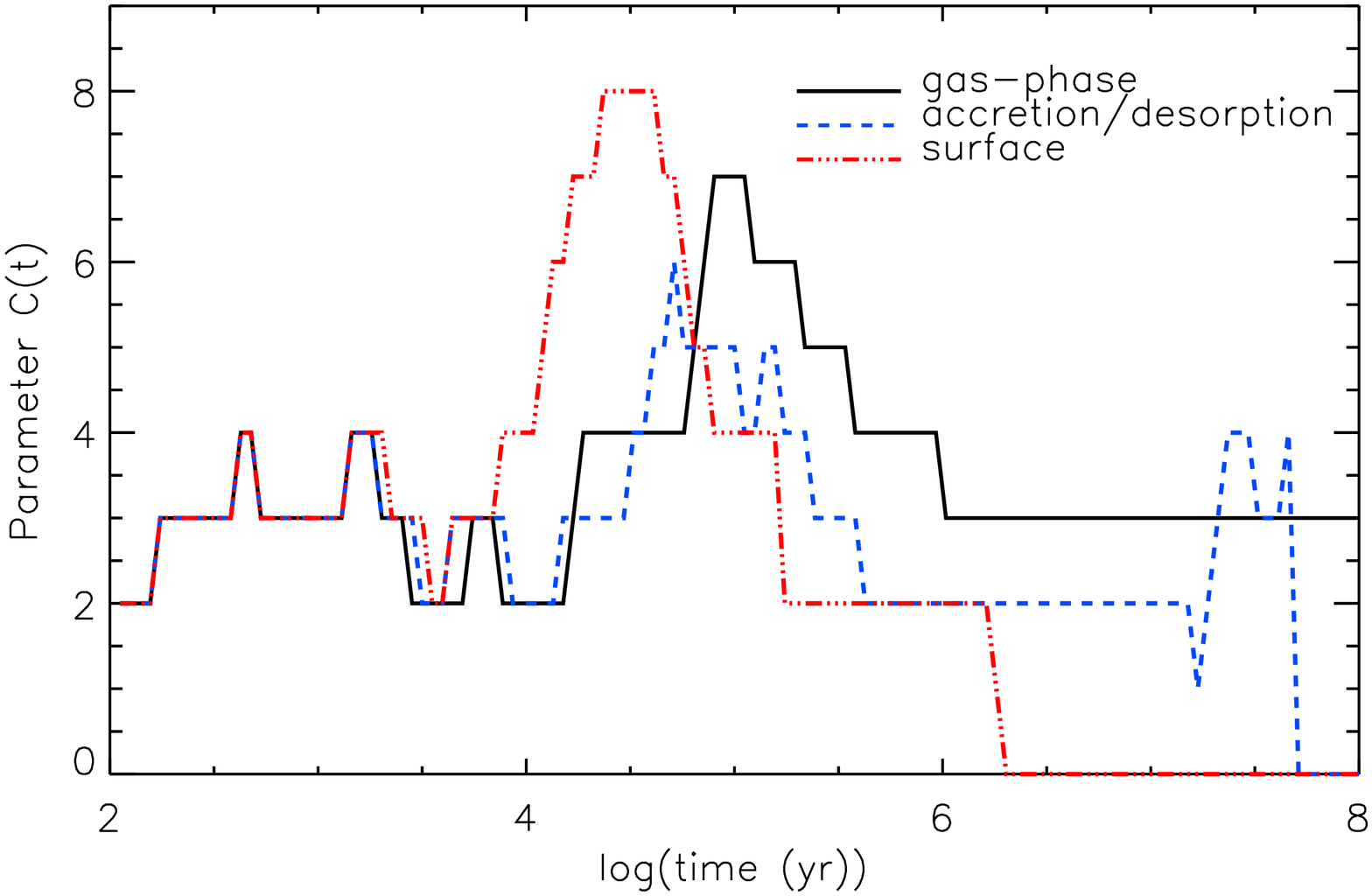}
	
	\includegraphics[width=5cm,angle=90]{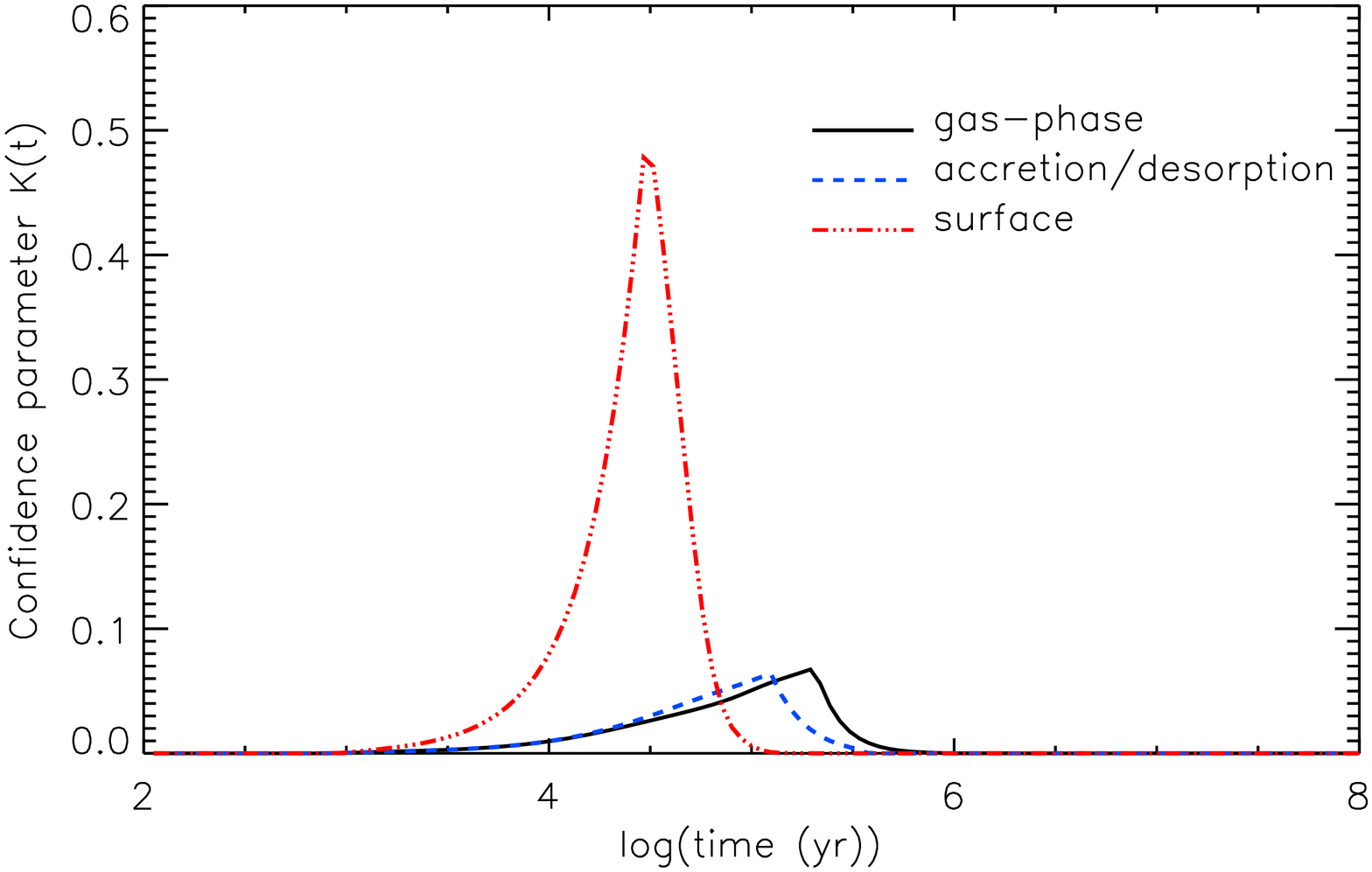}
	
	\caption{Topmost panels: abundance profiles relative to H$_2$ as a
		function of time. 
		The box on every panel corresponds to the observational
		values for IRDC321.73 $\pm$ one order of magnitude.
		Middle panel: Time dependence of the parameter $C(t)$ for the three different networks.
		Lowest panel: Time dependence of the confidence parameter $K(t)$
		for the three different networks. 
		 On the topmost panels, the solid black line corresponds to the gas-phase network, dashed  line to 
		the accretion/desorption network, and dash-dotted line to the surface network.
		(A color version of this figure is available in the online journal.)}
\label{figure:25K_105}
\end{figure}

The topmost part of Figure~\ref{figure:25K_105} shows observational abundances together with modeled abundances
at $T = 25$~K and $n(H_{2}) = 10^5$ cm$^{-3}$.
Under these conditions, the gas-phase and accretion/desorption networks show similar
abundance profiles, whereas the surface network gives different results.
Unlike the situation at 15 K, the gaseous CO abundance stays high for a considerable
amount of time for the accretion/desorption case, but not for the surface case.
Even here, the freeze-out starts somewhat later than at the lower temperature.
 As the freeze-out occurs, the
HCO$^+$  abundance once again undergoes a steep decline.
At the same time, the N$_2$H$^+$ abundance once again increases.
As can be seen in Figure 3, at  25~K only the surface network allows us to reproduce the observed N$_2$H$^+$ abundance at reasonable times.

The middle panel of Figure~\ref{figure:25K_105} shows that only the surface network yields
the highest value for $C(t)$ of 8,  which occurs from  2$\times$10$^4$ yr to 4$\times$10$^4$ yr.
    For the surface network, the peak value of $K(t)$, which is the highest of the three networks  by a considerable margin, occurs at $ 3 \times 10^{4}$~yr.
Modeled abundance values are presented in Table~\ref{table:abundances} for the surface network with $n = 10^{5}$ cm$^{-3}$ and $T = 25$~K at the  time when the confidence parameter $K(t)$ reaches its maximum.

As can be seen in Table~\ref{table:Ct},  the surface network  prefers
physical conditions in the range $T = 20 - 25$ K with $n = 10^5$ cm$^{-3}$, where the maximum
$C(t)=8$, although the fall-off at both lower and higher temperatures is only to $C(t) = 7$.
The $K(t)$ criterion yields a similar result.  Thus, unlike the cold case, the somewhat higher
temperature of IRDC321.73~-~1 is reproduced to some extent by  chemical modeling with the surface network.

We present  predictions for the abundance profiles of 12 species in Figure 4 that have not yet been observed.
The panels on the left of the figure correspond to the ``cold" case (15 K), while right panels correspond  
to the ``warm" case (25 K); in both
cases the surface network was used.  
In exploring differences between the results at the two temperatures,  
the sulfur-bearing species SO, 
SO$_{2}$ and C$_{2}$S stand out with 1-2 order-of-magnitude differences in abundance.  
Hence, observation of these species can reduce the temperature uncertainties deduced from model results,
 which is in agreement with previous work by  \citet{2004A&A...422..159W,2010A&A...517A..21W}.
The prediction of low fractional abundances for C$_{2}$S at both temperatures in the range 10$^{4-6}$ yr 
is in agreement with the observations of \citet{2008ApJ...678.1049S}.

\section{Discussion of the Chemistry}

In general, the calculated molecular abundances depend least on temperature for the gas-phase network and
more strongly for networks involving gas-grain interactions.  This can be seen by comparing results in
Figures~\ref{figure:gasph}, \ref{figure:accdes}, and \ref{figure:surface}, which show calculated molecular
profiles  for the observed species with the gas-phase, accretion/desorption, and surface networks, respectively.
The physical conditions depicted include temperatures of 10 K, 25 K, and 40 K, at H$_{2}$ densities of
10$^{5}$ cm$^{-3}$ and  10$^{6}$ cm$^{-3}$. These results supplement the molecular profiles already shown
in Figures~\ref{figure:15K_106} and \ref{figure:25K_105} for 15 K and 25 K at the lower density.
Given the much stronger sensitivity to temperature of surface processes,  
the result is not remarkable.

The chemical differences between the accretion/desorption and surface networks at 25~K 
are of particular interest.
The accretion/desorption network  produces abundance profiles
similar to  the gas-phase  rather than to the surface network.
This pattern occurs because the abundance of CO and, hence, HCO$^+$, does not decrease until 10$^{8}$ yr except with the surface network.
In the accretion/desorption network, at T=25 K,  the freeze-out of CO onto grain surfaces:
\begin{equation}
\rm CO +  Freeze out \rightarrow  grCO
\end{equation}
\noindent
is compensated by desorption back into the gas phase:
\begin{equation}
\rm grCO + Desorption \rightarrow  CO ,
\end{equation}
where gr stands for a species on the grain surface.
In the surface network,  on the other hand, when CO freezes out on the grain surface,
it does not desorb predominantly back into the gas phase.  Rather, it is destroyed more rapidly by reactions with other surface species; e.g.,
\begin{equation}
\rm grCO +  grOH \rightarrow  grCO_2 + grH,
\label{equ:gCOgOH}
\end{equation}
\begin{equation}
\rm grCO +  grS \rightarrow  grOCS,
\label{equ:gCOgS}
\end{equation}
\noindent
These two reactions are the most efficient processes destroying CO on grain surfaces in our model.   The reaction with sulfur has no barrier, while the reaction with OH has a small activation barrier of 80~K. This barrier significantly reduces its rate at 10 K, but has  little influence on the rate at 25 K.

At first glance, the low abundance of gas-phase CO in the model with surface reactions might totally explain the enhancement of N$_2$H$^+$, which can be seen in Figures~\ref{figure:25K_105} and \ref{figure:surface}.   The main destruction route of N$_2$H$^+$ is via the reaction
\begin{equation}
\rm N_2H^+ +  CO \rightarrow  HCO^+ + N_2,
\label{equ:N2H+dest}
\end{equation}
the importance of which has been discussed in the literature \citep{2007ARA&A..45..339B} 
for a variety of sources, ever since the initial work of \citet{1977ApJ...212...79S}.  
Via this reaction, a decrease in the gaseous CO abundance leads directly to an increase 
in the N$_2$H$^+$ abundance  as long as the production rate of the ion is maintained, a 
situation that need not occur if gaseous N$_{2}$ is also depleted.  Indeed, for low-mass 
starless cores, the abundance of N$_{2}$H$^{+}$ can be constant in abundance towards the 
center despite the loss of gaseous CO \citep{2004A&A...416..191T}.  Towards low-mass 
protostellar cores, there is some evidence for an increase in the abundance of N$_{2}$H$^{+}$ 
based on the study of the abundance ratio NH$_{3}$/N$_{2}$H$^{+}$ \citep{2002ApJ...572..238C}.  
The situation of intermediate-mass protostellar sources is more complex; although models with a 
constant abundance of N$_{2}$H$^{+}$ fit some of the data,  other data are fit with a depletion 
of this ion when gaseous CO is depleted \citep{2010A&A...518A..52A}.   
Finally, the case of N$_{2}$H$^{+}$ towards the massive protostellar cluster AFGL 5142 shows 
this ion and HCO$^+$ to be anti-correlated in the central core because, as the authors suggest, 
 the high temperature (70 K) allows the CO to evaporate from grain mantles \citep{2011A&A...525A.141B}.  
 
In our study of IRDCs, we find that the situation is also complex, since here reaction (\ref{equ:N2H+dest}) is not the sole cause of the enhancement of N$_{2}$H$^{+}$ at higher temperatures. 
If we look carefully at the expanded scale in Figure~\ref{figure:suface_one}  for 25~K, we notice that the
abundance of N$_2$H$^+$ starts to increase before the decrease of CO  for the surface network.  
To fully explain the initial increase in  N$_2$H$^+$, it is necessary to consider  its principal formation reaction:
\begin{equation}
\rm H_3^+ +  N_2 \rightarrow  N_2H^+ + H_2.
\label{equ:N2H+}
\end{equation}
Figure~\ref{figure:suface_one}  shows that the N$_2$ abundance profiles are quite similar for the surface and accretion/desorption networks
whereas the abundance of H$_3^+$ becomes higher after 10$^{3}$ yr if we use the complete surface network.  This difference can be explained in the follow manner.
At very early times  (10$^{3}$ yr $<$  t $<$ 10$^{4}$ yr) H$_3^+$ reacts primarily with O via  

\begin{equation}
\rm H_3^+ +  O \rightarrow  OH^+ + H_2,
\label{equ:H3O}
\end{equation}
and only later  primarily with CO:
\begin{equation}
\rm H_3^+ +  CO \rightarrow  HCO^+ + H_2.
\label{equ:H3CO}
\end{equation}
Just as with CO, O atoms on the surface react rather than desorb in the surface network.  Thus, accretion turns out to be a destruction process for gaseous O as well as gaseous CO.   A reduction in gas-phase atomic oxygen leads to an increase in the abundance of
H$_3^+$ at early times and thus to  an early increase in the N$_2$H$^+$ abundance through reaction (\ref{equ:N2H+}) for the surface network.  Later, at $\sim$ 10$^{4.5}$ yr, the N$_2$H$^+$ abundance is further increased due to the rapid drop in the gas-phase CO abundance.

With the surface network at 15 K, the N$_2$H$^+$ abundance, and to a lesser degree the NH$_3$ abundance, differs somewhat
from what we see at 25 K in Figure~\ref{figure:suface_one}.
At the lower temperature, surface CO reacts more slowly, 
so that desorption, although slower than at 25 K, is sufficiently competitive  to maintain a 
significant abundance of gaseous CO.  
On the other hand,  a large amount of N$_2$ stays on the grain surface for long periods of time at 15 K, 
while at 25 K there is enough thermal energy to desorb it into the gas-phase.  
Taking into account reaction (\ref{equ:N2H+}),  this leads to a somewhat lower N$_2$H$^+$ abundance at 15 K than at 25 K  throughout much of the time of the calculations.  Although the distinction is minimal at the time of maximum $K(t)$, as shown in Table \ref{table:abundances}, it can range up to an order of magnitude, as shown in Figure~\ref{figure:suface_one}.

 Now let us compare the abundance of N$_{2}$H$^{+}$ among different types of sources.
An interplay among the formation of N$_2$H$^+$ via process 
(\ref{equ:N2H+}), its destruction through reaction (\ref{equ:N2H+dest}), 
and destruction of CO through grain-surface processes (\ref{equ:gCOgOH}) 
and (\ref{equ:gCOgS}) in the surface network allows us to qualitatively explain the systematic 
differences in the N$_2$H$^+$ abundance among low-mass starless cores, 
high-mass protostellar objects and IRDCs noticed in \citet{2011A&A...527A..88V}. 
In this investigation,   higher N$_2$H$^+$ abundances were determined for IRDCs in comparison with 
both low-mass starless cores and high-mass protostellar objects (HMPOs). 
In our explanation, we take into account that typical temperatures in these three 
classes of objects are different. Low-mass starless cores have typical
temperatures of $\sim$10~K \citep[e.g.][]{2004A&A...416..191T}, some IRDCs reach temperatures of about 15~K~--~25~K \citep{2011A&A...527A..88V}, 
and HMPO temperatures are higher than 30~K \citep{2007prpl.conf..165B}. At 10~K,  N$_2$, 
the major precursor of N$_2$H$^+$, is partially frozen onto the dust particles,
but some sufficient fraction of  it still resides in the gas phase, making possible
the formation of a moderate amount of gaseous N$_2$H$^+$
 in low-mass starless cores, as seen in observations. 
HMPOs, on the other hand,  are warm enough to maintain a high gas-phase abundance of N$_2$, but here the
CO gaseous abundance also stays high   because of the rapid desorption (see Figure~\ref{figure:surface}), so that 
CO can react with N$_2$H$^+$ via reaction (\ref{equ:N2H+dest})  to suppress the ionic abundance. 
Finally, warm IRDC321.73-1 represents the most comfortable environment for N$_2$H$^+$ because 
it is warm enough (T~$\sim$~25~K) to keep all N$_2$ in the gas phase, while the majority of CO 
is  removed from the gas via accretion followed by reactions (\ref{equ:gCOgOH}) and (\ref{equ:gCOgS}). 
These processes lead to an N$_2$H$^+$ abundance that is  on average higher in IRDCs than in low-mass starless cores and HMPOs. 
Also, it explains the higher observed abundance of N$_2$H$^+$ in warm IRDC321.71-1 than 
in cold IRDC013.90-1,  the chemistry of  which is much more similar to the chemistry of low-mass starless cores.

The fact that the abundances of gas-phase species such as CO, HCO$^+$ and N$_2$H$^+$ in the surface model of warm IRDC321.73-1 are controlled by the
grain-surface chemistry of CO  leads us to the interesting conclusion that
grain-surface chemistry can be investigated not only via direct
observations of ices or complex organic molecules, but also by means
of careful observations of the chemistry of simple molecular ions.   This
is possible when the temperature of dust in a source stays in the
range between 20~K and 30~K,  which ensures the simultaneous participation
of reactive gas-phase molecules such as CO both in the gas-phase and on grains.
 Such a warm medium is not typical for
low-mass pre-stellar objects, but seems to be present in some
objects that represent the earliest stages of  high-mass star
formation. Therefore, warm infrared dark clouds may serve as an unusual and
very useful laboratory that makes possible the investigation of
grain-surface chemistry through the observations of simple gas phase
species.

\section{Summary and Conclusions}

Infrared dark clouds represent a new area for chemical study in which the temperatures and densities are somewhat higher than in low-mass cold prestellar cores.   The recent study of a number of gas-phase molecules via single-dish observations allows us to compare observed molecular abundances with the results of chemical simulations and to determine in some detail how the chemistry in IRDCs compares with the chemistry in low-mass sources.   To study the chemistry, we chose the infrared dark clouds IRDC013.90-1 and IRDC321.73-1, which are relatively well studied observationally and  represent different temperature conditions for IRDCs: $\approx 15$~K for the former and $\approx 25$~K for the latter.   Given the large beam of the single-dish observations, we chose a simple 0-D model with  time-independent physical conditions.
The 0-D model  was used f with three different networks: the surface network, which contains the complete OSU gas-grain network,  the gas-phase network, in which only gas-phase processes occur except for the surface formation of H$_{2}$ and ion-grain reactions, and the accretion/desorption network, in which
accretion and desorption are added to the gas-phase network.

We used two time-dependent criteria to determine the goodness-of-fit of the model results to the data -- the simple criterion of counting the number of species for which theory and observation agree to within an order of magnitude, and the more subtle confidence parameter approach.   With these criteria, we found for IRDC321.73~-~1 that  the best fit occurs at times below 10$^{5}$ yr with the surface network at 20-25 K and $n = 10^5$ cm$^{-3}$, in good agreement with the ammonia kinetic temperature.  This best fit, however, is not sharply peaked in temperature, and fits for temperatures lower than or higher than the 20-25 K range do almost as well.
For IRDC013.90-1, it is even more difficult to determine the optimum temperature from best fits with our two criteria.  
All we can say is that the physical conditions based on observations -- $T$ = 15 K and $n = 10^5$ cm$^{-3}$ -- are among 
the more reasonable choices with the surface network.  Analyzing predicted molecular profiles for species not included in our study, we conclude
that additional observations of sulfur-bearing species can help 
to sharpen the temperature range of the best fits.

 We looked at model results for the three networks in detail over a temperature range from 10 K to 40 K at densities of $n = 10^5$ cm$^{-3}$ and $n = 10^6$ cm$^{-3}$.  For these ranges, we can see that
 only in the case of the gas-phase network  do
 temperature and density have no large influence on molecular abundances.
 As soon as we take gas-grain interactions into account,  the
 temperature dependence can become quite large, especially at later times.
Our study revealed a limited temperature range of 20 - 30~K, where grain-surface chemistry becomes important for gas-phase abundances of certain species such as CO, HCO$^+$, N$_2$H$^+$.   
 Hence, those IRDCs that have temperatures in the  20 - 30 K range can be a good laboratory
 to study the impact of surface reactions on the abundances of simple gas-phase species.  
 Moreover, the interplay of gas-phase and surface chemistry can distinguish low-mass prestellar 
 cores, IRDCs and higher-mass protostellar objects because their different temperatures 
 lead to different abundances of the ion N$_{2}$H$^{+}$.   
 The chemistry can also explain the differing abundances of this ion  when comparing colder and warmer IRDCs.
  The highest N$_{2}$H$^{+}$ fractional abundance occurs in warm IRDCs (up to 10$^{-8}$), a feature that is reproduced quantitatively by our simple 
surface model.  

 The success of the 0-D model in reproducing single-dish observations of 8 species
 confirms that it is sufficient to use simple  physical models for explaining this type of data.
 To  build 1-D models with some temperature and density structure
 requires extending the number of species including tracers of more dense gas
 (N$_2$D$^+$ and DCO$^+$) and less dense gas (CO).
 To build and justify more complicated models,
 including contraction and heating, we need high-resolution interferometric data \citep[e.g.,][]{2011ApJ...739L..16B}.
 Interferometers such as  ALMA will allow us to achieve the same spatial resolution
 for distant IRDCs as we have now with large single-dish telescopes for  closer low-mass clouds.

\begin{table}
\caption{Observational and modeled abundances of molecules with respect to H$_2$ in IRDC321.73-1 and IRDC013.90-1} %title of Table
\label{table:abundances} 
\begin{center}     % is used to refer this table in the text
%\centering                          % used for centering table
\begin{tabular}{l c c c c }
\hline            % inserts double horizontal lines
%\noalign{\smallskip}
    &  \multicolumn{2}{c}{IRDC013.90-1} &  \multicolumn{2}{c}{IRDC321.73-1} \\
%\noalign{\smallskip}
\hline
%\noalign{\smallskip}
Species & Observation & Model  & Observation & Model \\
                &                          &  8$\times$10$^4$ yr &  & 3$\times$10$^4$ yr   \\
%\noalign{\smallskip}
\hline                        % inserts single horizontal line
%\noalign{\smallskip}

 N$_2$H$^+$  &   6.4(-10)  &  1.3(-09) &   3.9(-09)  &  1.4(-09)      \\
 HCCCN       &   1.5(-10)  &  9.0(-11) &   5.1(-10)  &  7.1(-10) 	  \\
 HNC         &   4.0(-10)  &  1.3(-09) &   1.8(-09)  &  2.1(-09) 	  \\
 HCO$^+$     &   3.8(-09)  &  1.6(-09) &   1.3(-08)  &  3.1(-09)	 \\
 HCN         &   4.2(-10)  &  1.4(-09) &   3.1(-09)  &  2.0(-09) \\
 C$_2$H      &   2.5(-09)  &  3.5(-09) &   1.4(-08)  &  3.3(-09)	 \\
 NH$_3$      &   4.8(-08)  &  1.4(-08) &   8.0(-08)  &  2.3(-08) 	 \\
 CS          &   --	   &  --       &    3.9(-09)  &  7.4(-10)    \\

%\noalign{\smallskip}
%\hline
%\noalign{\smallskip}  
\hline                              %inserts single line
\end{tabular}
\end{center}
\tablecomments{Modeled values  are obtained at time of maximum $K(t)$ with the surface network. For IRDC321.73-1,  $T$ =25 K, $n=10^5$ cm;$^{-3}$; for IRDC013.90-1: $T$ =15 K, $n=10^5$ cm$^{-3}$.} 
\tablecomments{a(b) denotes $ a \times 10^b$.}
\end{table}

\begin{table}
\caption{Initial abundances of species with respect to $n_{\rm H}=n({\rm H})+2n({\rm H_{2}})$\tablenotemark{a} } %title of Table
\label{table:initial}      % is used to refer this table in the text
%\centering                          % used for centering table
\begin{tabular}{l c}
\hline            % inserts double horizontal lines
\noalign{\smallskip}
Species & Abundance  \\
\noalign{\smallskip}
\hline                        % inserts single horizontal line
\noalign{\smallskip}

H$_2$           &  0.50  \\
He           &  1.40(-1)  \\
N            &  2.14(-5)  \\
O            &  1.76(-5)  \\
C$^+$        &  7.30(-5)  \\
S$^+$        &  8.00(-8)  \\
Si$^+$       &  8.00(-9)  \\
Fe$^+$       &  3.00(-9)  \\
Na$^+$       &  2.00(-9)  \\
Mg$^+$       &  7.00(-9)  \\
Cl$^+$       &  1.00(-9)  \\
P$^+$        &  2.00(-10) \\
F$^+$        &  6.68(-9)  \\

\noalign{\smallskip}
\hline                                   %inserts single line
\end{tabular}
\tablecomments{a(b) denotes $ a \times 10^b$.}
\tablenotetext{a}{Set EA1 from \citet{2008ApJ...680..371W}.}	
\end{table}

\begin{table}

\caption{Parameters of the model} %title of Table
\label{table:initial_cond}      % is used to refer this table in the text
%\centering                          % used for centering table
\begin{tabular}{r c}
\hline           % inserts double horizontal lines
\noalign{\smallskip}
Parameter & Value  \\
\noalign{\smallskip}
\hline                        % inserts single horizontal line
\noalign{\smallskip}

cosmic ray ionization rate                &  1.3(-17) s$^{-1}$ \\
Grain size                                &  0.1 $\mu$m  \\
Grain density                             &  3.0 g/cm$^3$  \\
Diffusion/desorption energy ratio        &  0.5  \\
Gas-to-dust mass ratio                    &  100  \\
Extinction                                &  10   \\
Surface site density                      &  1.5(15) cm$^{-2}$ \\
Thermal desorption                        &  on  \\
Cosmic ray desorption                     &  on  \\
Photodesorption                          &  off  \\
Tunneling   through diffusion barriers                              &  off  \\
Tunneling through reaction barriers       &  on  \\
Thermal hopping                           &  on  \\
\noalign{\smallskip}
\hline                                   %inserts single line
\end{tabular}
\tablecomments{	a(b) denotes $ a \times 10^b$.}
	
\end{table}

\begin{table}
\caption{Maximum values of C(t) and K(t) for each network. } %title of Table
\label{table:Ct}      % is used to refer this table in the text
%\centering                          % used for centering table
\begin{tabular}{l c c c c c}
\hline           % inserts double horizontal lines
\noalign{\smallskip}
n / T & 10 K & 15 K & 20 K & 25 K & 30 K \\
\noalign{\smallskip}
\hline                        % inserts single horizontal line
\noalign{\smallskip}

 \multicolumn{6}{c}{C(t) for IRDC013.90-1} \\

10$^5$ cm$^{-3}$ & 6, {\bf 7}, {\bf 7} & 6, {\bf 7}, {\bf 7} & 6, 7p, {\bf 7} & 6, 6, {\bf 7} & 6, 6, {\bf 7} \\

10$^6$ cm$^{-3}$ & 4, 2, 6 & 5, 3, 6 & 5, 5, {\bf 7}  & 5, 4, {\bf 7} & 5, 4, 6p \\

 \multicolumn{6}{c}{K(t) for IRDC013.90-1} \\

10$^5$ cm$^{-3}$ & 0.3, 0.4, 0.5  & 0.3, 0.4, {\bf 0.6}   & 0.3, 0.4, 0.4  & 0.3, 0.3, {\bf 0.6}     & 0.3, 0.3, 0.4 \\

 10$^6$ cm$^{-3}$ & 0.03, 0.1, 0.1 & 0.03, 0.08, 0.1 & 0.03, 0.1, 0.5 & 0.03, 0.05, 0.4   & 0.03, 0.1, 0.2 \\

 \multicolumn{6}{c}{C(t) for IRDC321.73-1} \\

 10$^5$ cm$^{-3}$ & 7, 4, 6  & 7, 6p, 7 & 7, 7, {\bf 8}  & 7, 6p, {\bf 8}  & 7, 7p, 7 \\

 10$^6$ cm$^{-3}$ & 5, 4, 6p & 4, 3, 5p & 4, 3, 7p & 4, 4, 5   & 5, 6p, 6p \\

 \multicolumn{6}{c}{K(t) for IRDC321.73-1} \\

10$^5$ cm$^{-3}$ & 0.07, 0.1, 0.3    & 0.07, 0.1, 0.4 & 0.06, 0.2, {\bf 0.5}  & 0.06, 0.06, {\bf 0.5}  & 0.06, 0.1, 0.2 \\

10$^6$ cm$^{-3}$ & 0.003, 0.02, 0.05 & 0.004, 0.01, 0.06 &0.004, 0.04, 0.2 & 0.004, 0.01, 0.2   & 0.004, 0.1, 0.05 \\

\noalign{\smallskip}
\hline                                   %inserts single line
\end{tabular}
\tablecomments{First number corresponds to the gas-phase network, second to the accretion/desorption network,
		and third to the surface network. The index ``p'' signifies that $C(t)$ reaches its maximum at only a brief interval.
		Maximum $C(t)$ and $ K(t)$ values are in bold font.}
\end{table}

%__________________________________________________________________

\acknowledgements

EH wishes to  acknowledge the support
of the National Science Foundation for his astrochemistry program, and his program in chemical kinetics through the Center for the
Chemistry of the Universe. He also acknowledges support from the NASA Exobiology and Evolutionary Biology program through a subcontract from Rensselaer Polytechnic Institute.

This research has made use of the NASA/ IPAC Infrared Science Archive, which is operated by
the Jet Propulsion Laboratory, California Institute of Technology, under contract with the
National Aeronautics and Space Administration.

\bibliographystyle{apj}
\bibliography{apj-jour,modeling23032012}

\begin{figure}
	\includegraphics[width=6cm,angle=90]{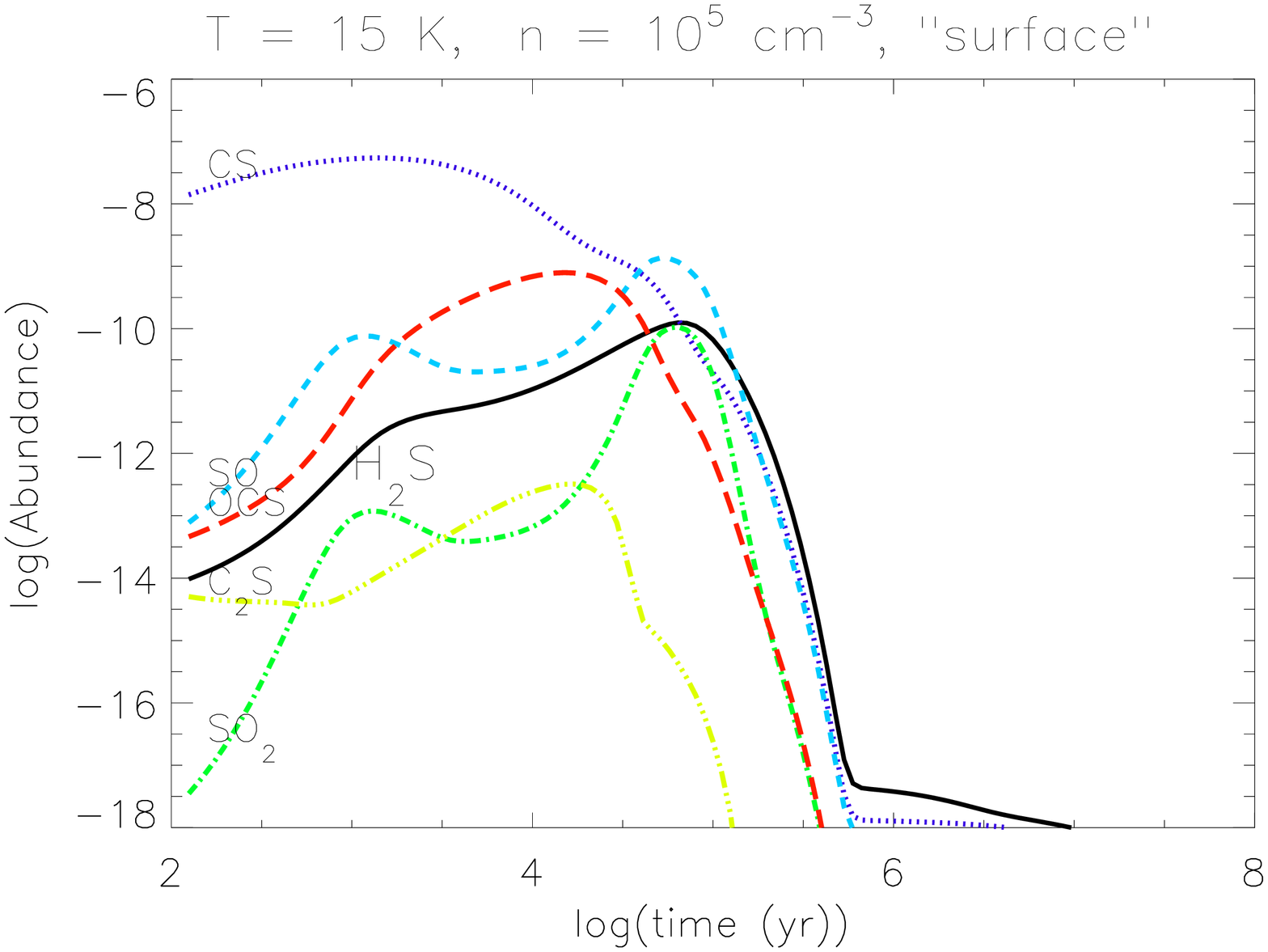}
	\includegraphics[width=6cm,angle=90]{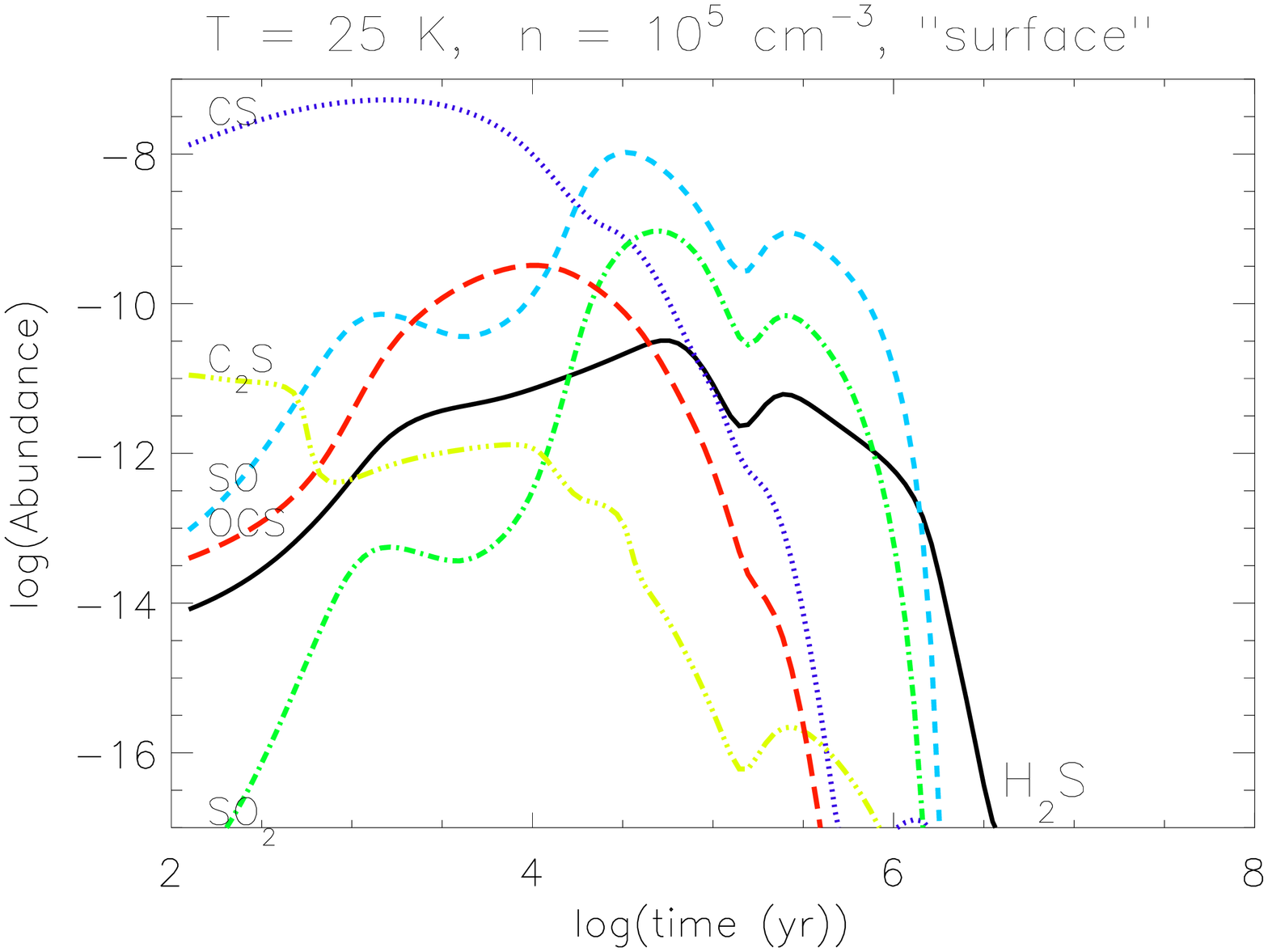}	
	
	\medskip
	\medskip
	\medskip
	\includegraphics[width=6cm,angle=90]{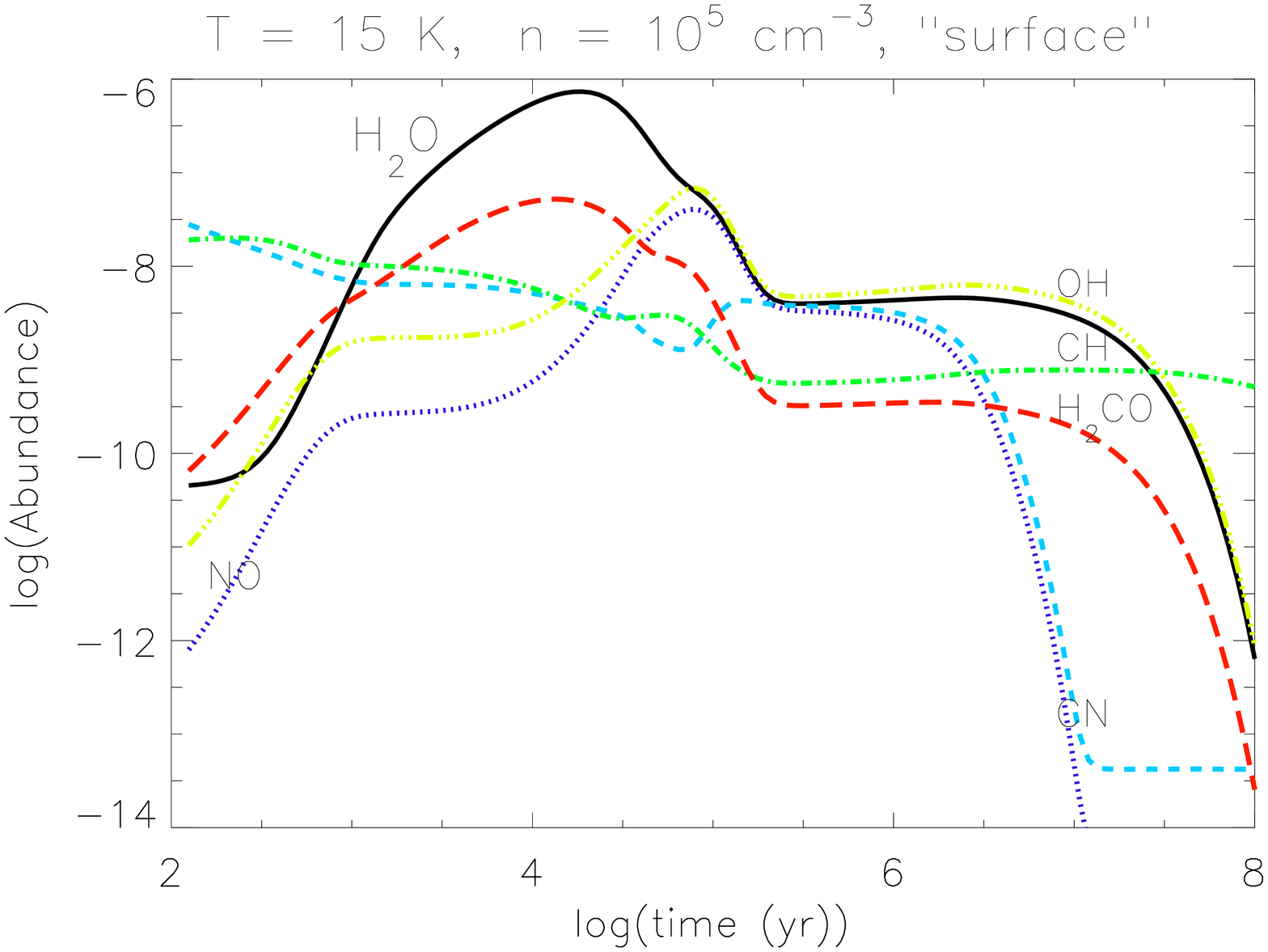}
	\includegraphics[width=6cm,angle=90]{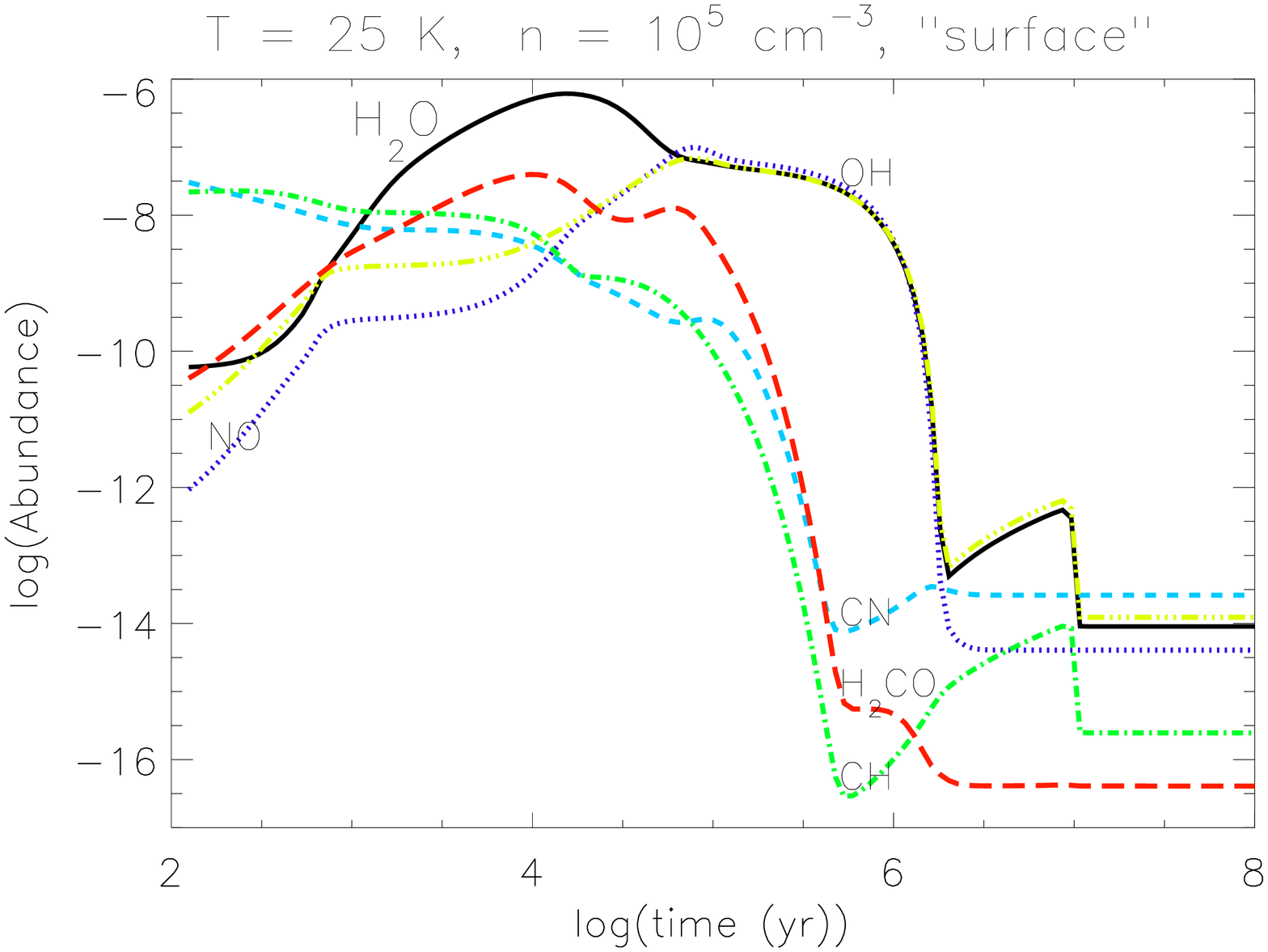}	
	\caption{Profiles of additional 12 species relative to H$_2$ as a function of time.
		Upper and lower left panels: $T$ = 15 K, $n$ =10$^5$ cm$^{-3}$, surface network.
		Upper and lower right panels: $T$ = 25 K, $n$ =10$^5$ cm$^{-3}$, surface network.
		(A color version of this figure is available in the online journal.)}
\label{figure:prediction}
\end{figure}


%% in parentheses, after the keyword, but they will not be verified.

{\it Facilities:} \facility{Nickel}, \facility{HST (STIS)}, \facility{CXO (ASIS)}.



\clearpage

%%%%%%%%%%%%%%%%%%%%%%%%%%%%%%%%%%%%%%%%%%%%%%%%%%%%%%%%%%%%%%%%

\begin{figure}
	\centering
	\includegraphics[width=7cm,angle=90]{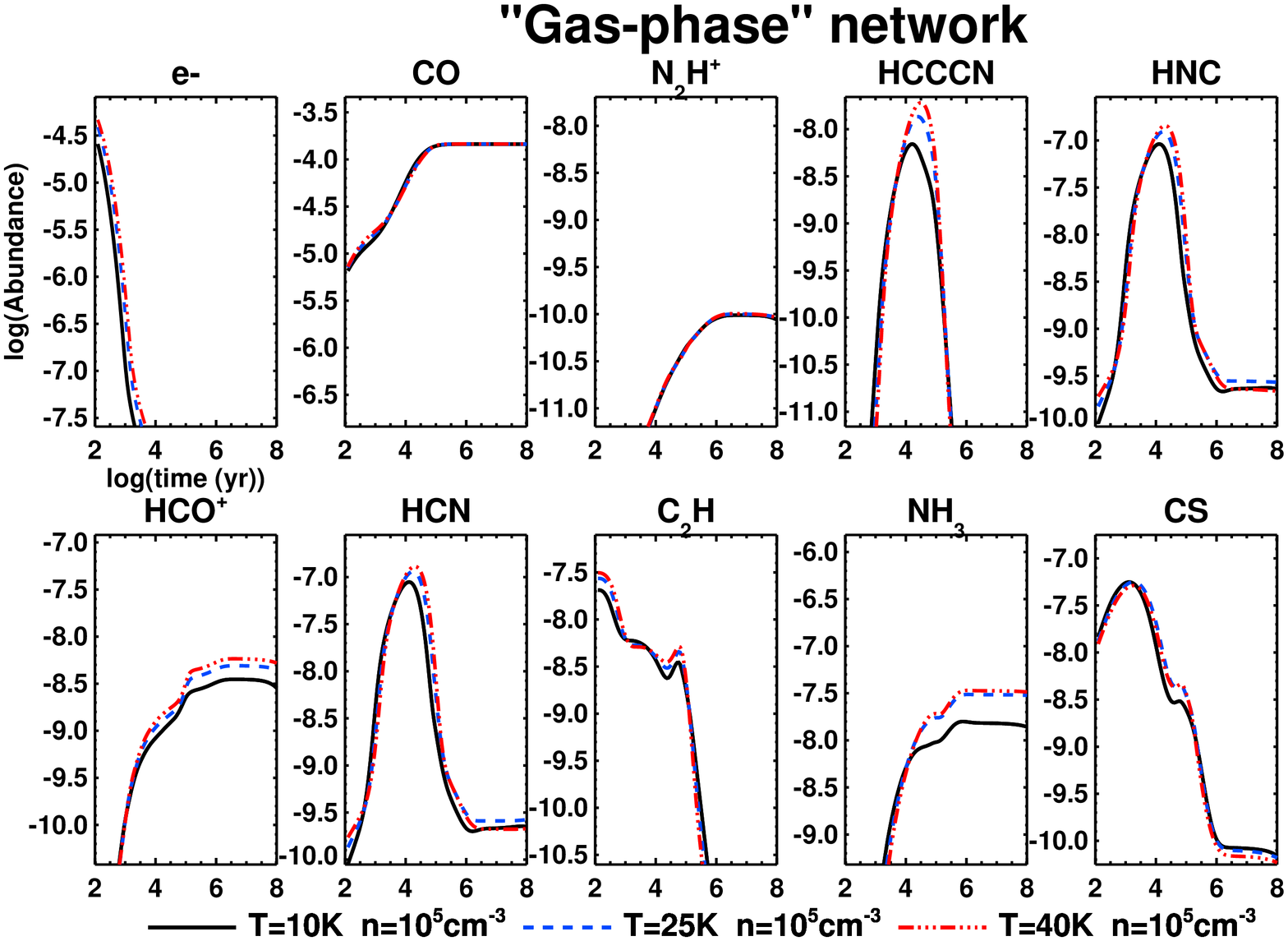}
		
	\medskip
	\medskip
	\medskip
	\medskip
	\includegraphics[width=7cm,angle=90]{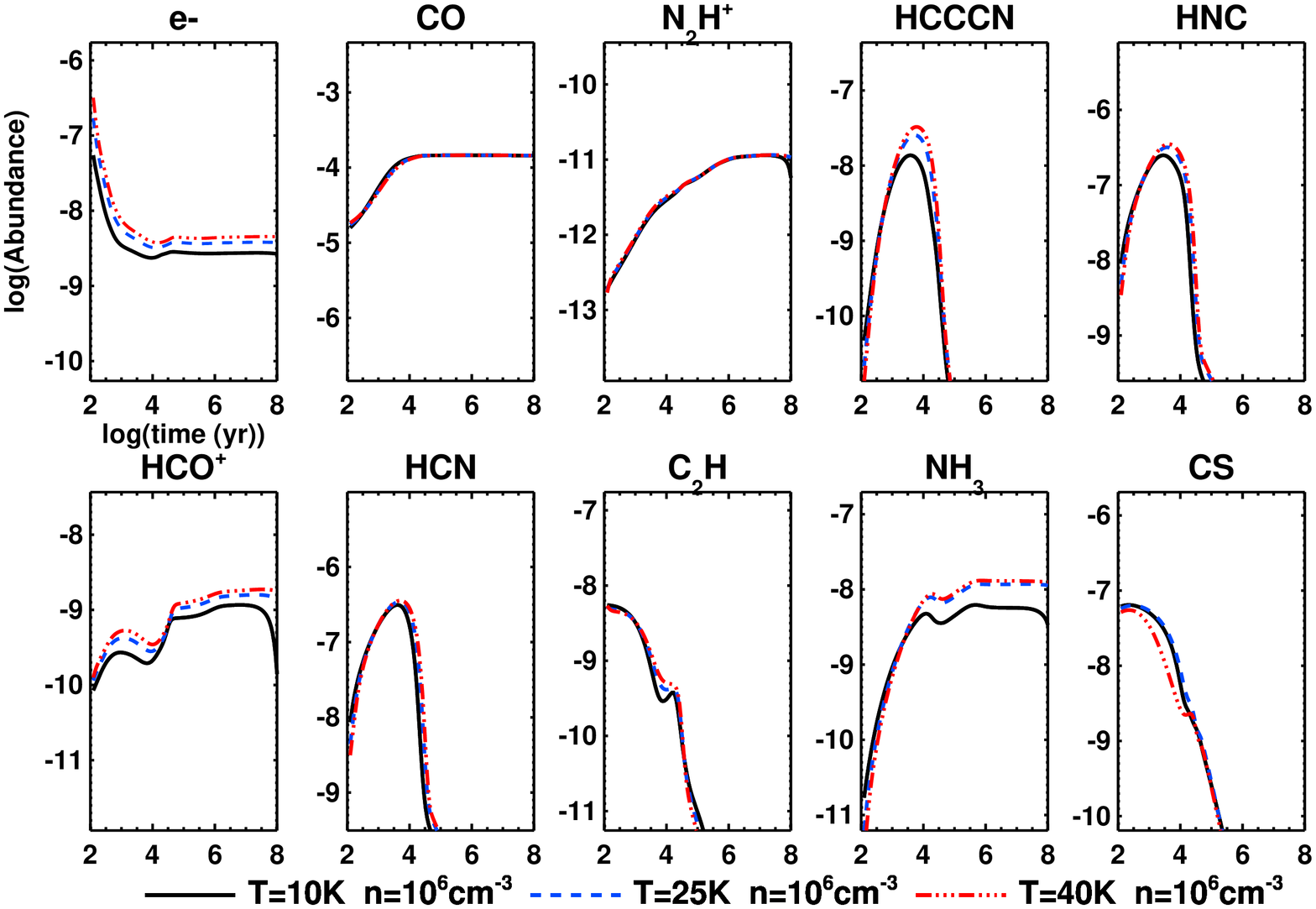}
	\caption{Abundance profiles relative to H$_{2}$ as a
		function of time obtained with the gas-phase network.
		Upper panel:  solid  line corresponds to
		T=10 K, n=10$^5$ cm$^{-3}$, dashed line  to
		T=25 K, n=10$^5$ cm$^{-3}$, dash-dotted  line  to
		T=40 K, n=10$^5$ cm$^{-3}$.
		Lower panel:  solid line corresponds to
		T=10 K, n=10$^6$ cm$^{-3}$, dashed line  to
		T=25 K, n=10$^6$ cm$^{-3}$, dash-dotted  line  to
		T=40 K, n=10$^6$ cm$^{-3}$.  (A color version of this figure is available in the online journal.)}
\label{figure:gasph}
\end{figure}

\begin{figure}
	\centering
	\includegraphics[width=7cm,angle=90]{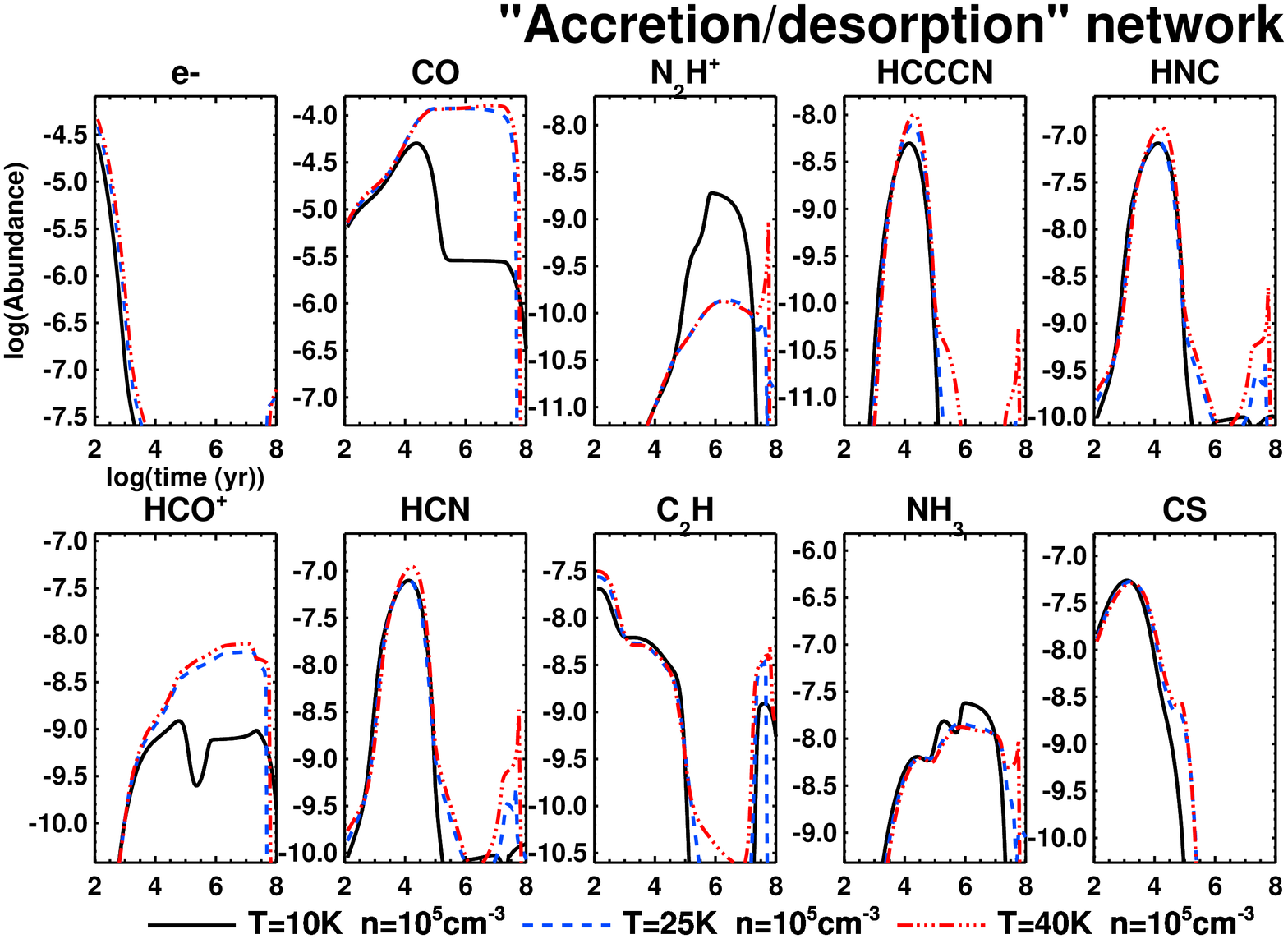}	
	
	\medskip
	\medskip
	\medskip
	\medskip
	\includegraphics[width=7cm,angle=90]{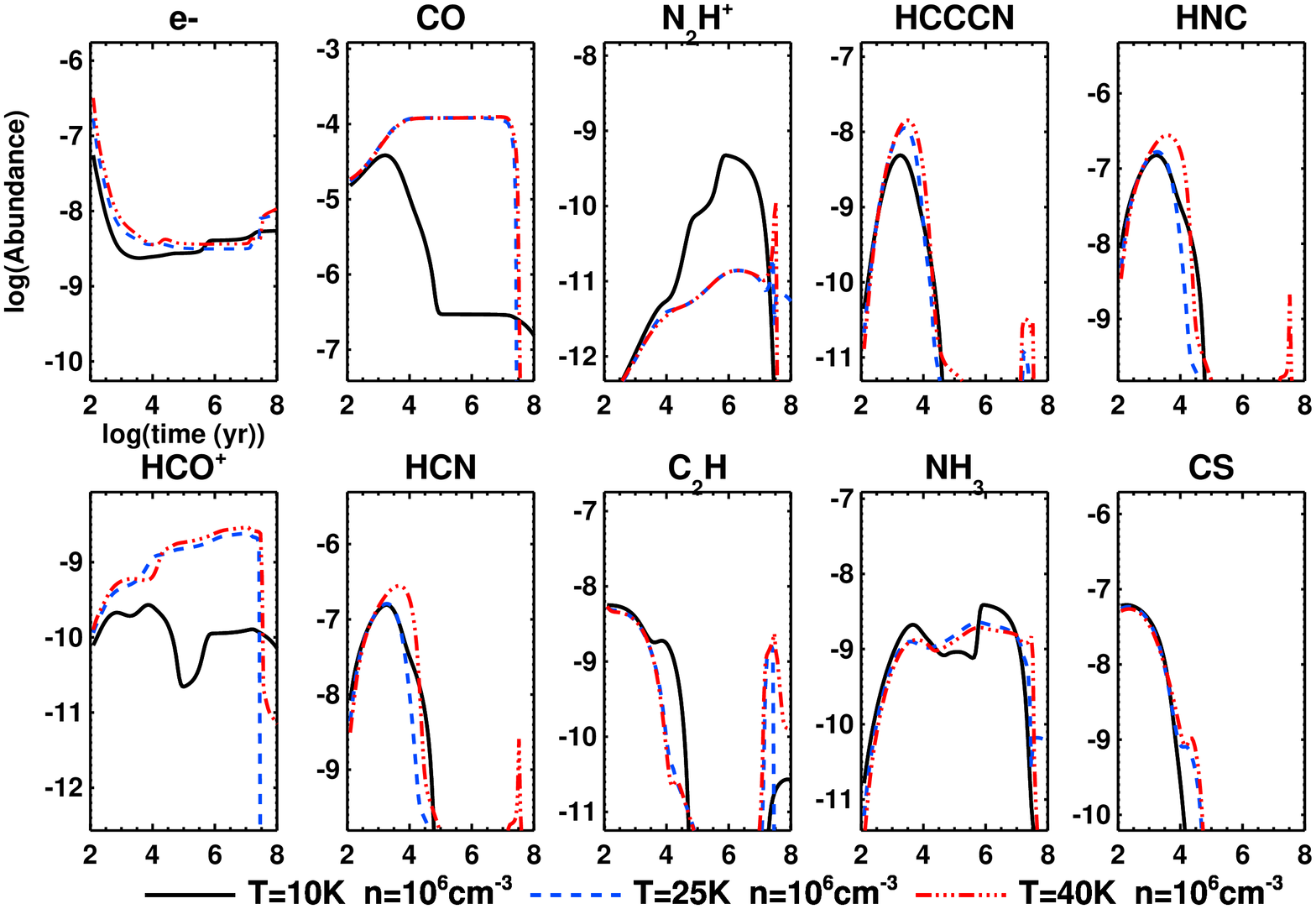}
	\caption{Abundance profiles relative to H$_2$ as a
		function of time  obtained with the accretion/desorption network.
		Upper panel:  solid line corresponds to
		T=10 K, n=10$^5$ cm$^{-3}$, dashed  line  to
		T=25 K, n=10$^5$ cm$^{-3}$, dash-dotted line  to
		T=40 K, n=10$^5$ cm$^{-3}$.
		Lower panel:  solidline corresponds to
		T=10 K, n=10$^6$ cm$^{-3}$, dashed line to
		T=25 K, n=10$^6$ cm$^{-3}$, dash-dotted  line  to
		T=40 K, n=10$^6$ cm$^{-3}$.  (A color version of this figure is available in the online journal.)}
		\label{figure:accdes}
\end{figure}

\begin{figure}
	\centering
	\includegraphics[width=7cm,angle=90]{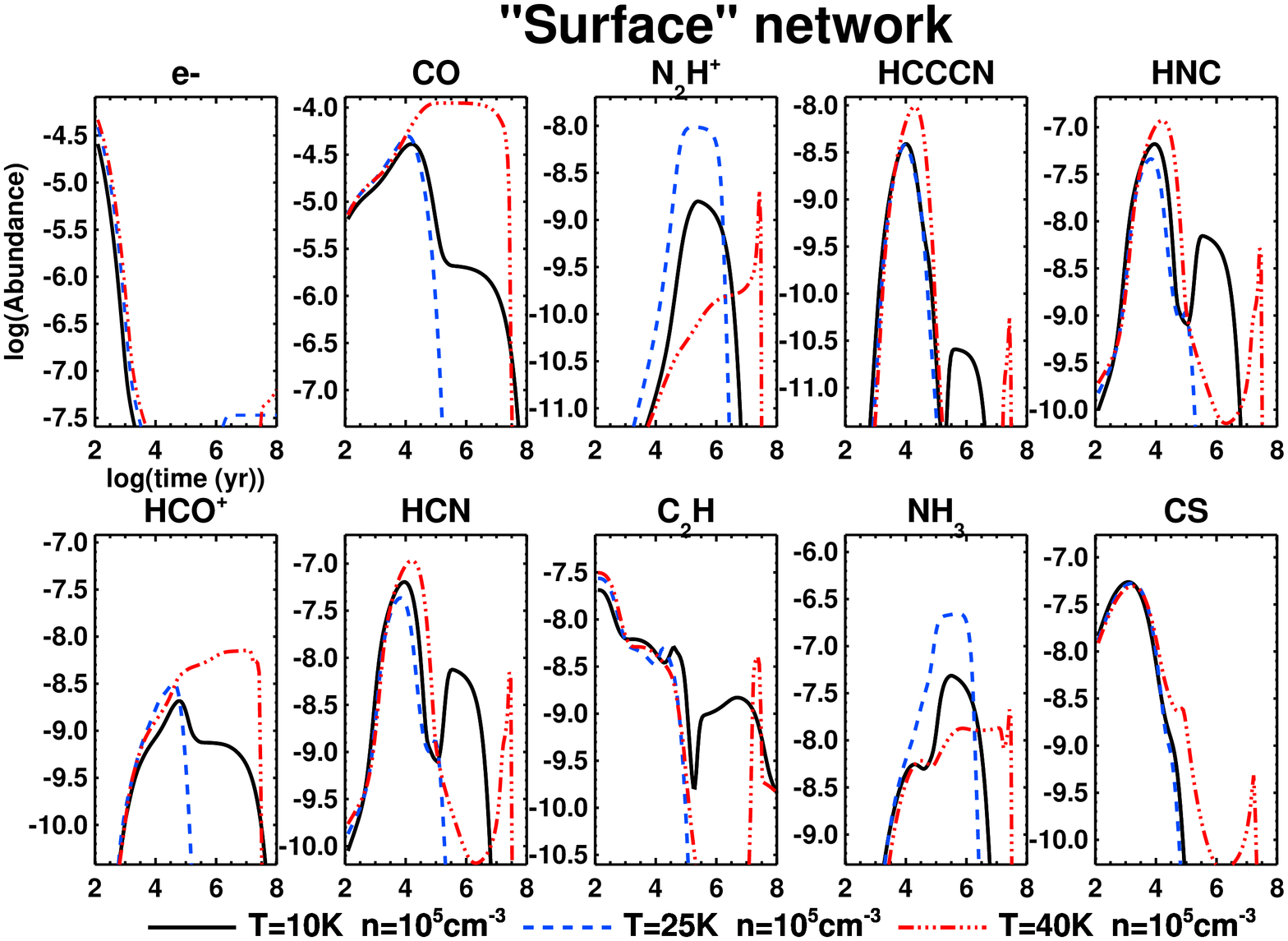}
		
	\medskip
	\medskip
	\medskip
	\medskip
	\includegraphics[width=7cm,angle=90]{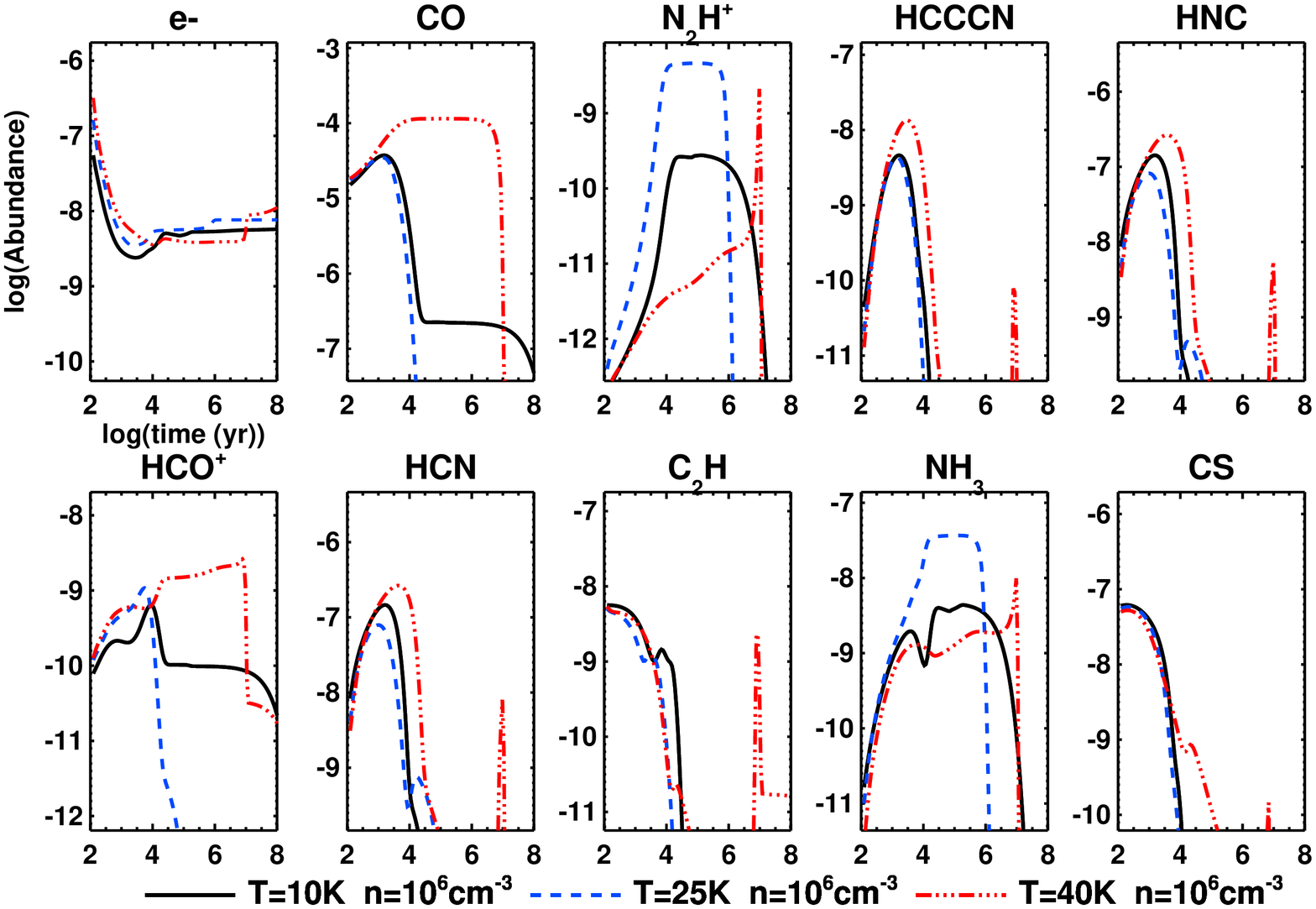}				
	\caption{Abundance profiles relative to H$_2$ as a
		function of time. Results are obtained by using the "surface" network.
		Upper panel:  solid  line corresponds to
		T=10 K, n=10$^5$ cm$^{-3}$, dashed  line  to
		T=25 K, n=10$^5$ cm$^{-3}$, dash-dotted line  to
		T=40 K, n=10$^5$ cm$^{-3}$.
		Lower panel:  solid  line corresponds to
		T=10 K, n=10$^6$ cm$^{-3}$, dashed line  to
		T=25 K, n=10$^6$ cm$^{-3}$, dash-dotted line  to
		T=40 K, n=10$^6$ cm$^{-3}$.  (A color version of this figure is available in the online journal.)}
\label{figure:surface}
\end{figure}

\clearpage

\begin{figure}
	\includegraphics[width=6cm,angle=90]{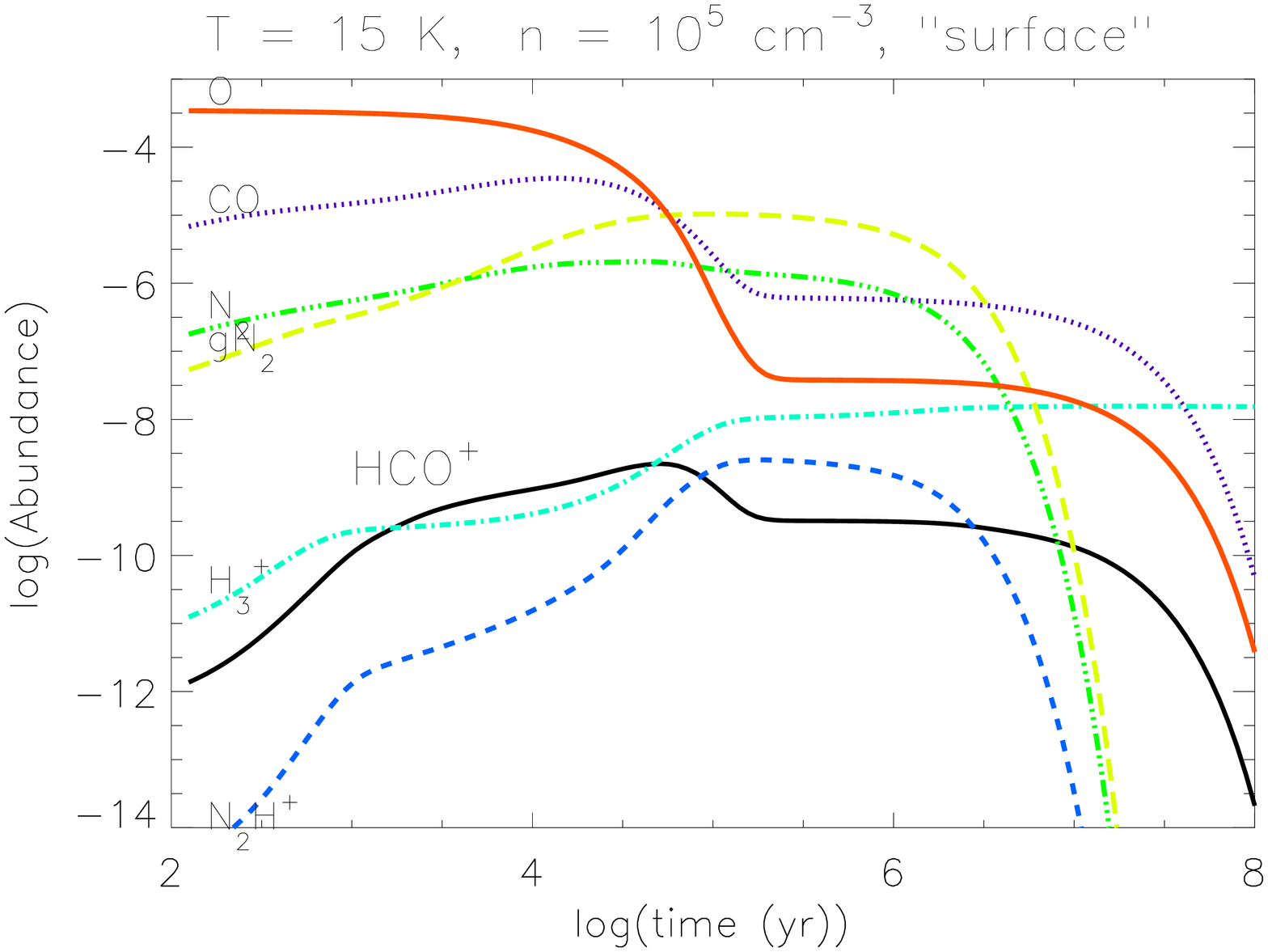}
	\includegraphics[width=6cm,angle=90]{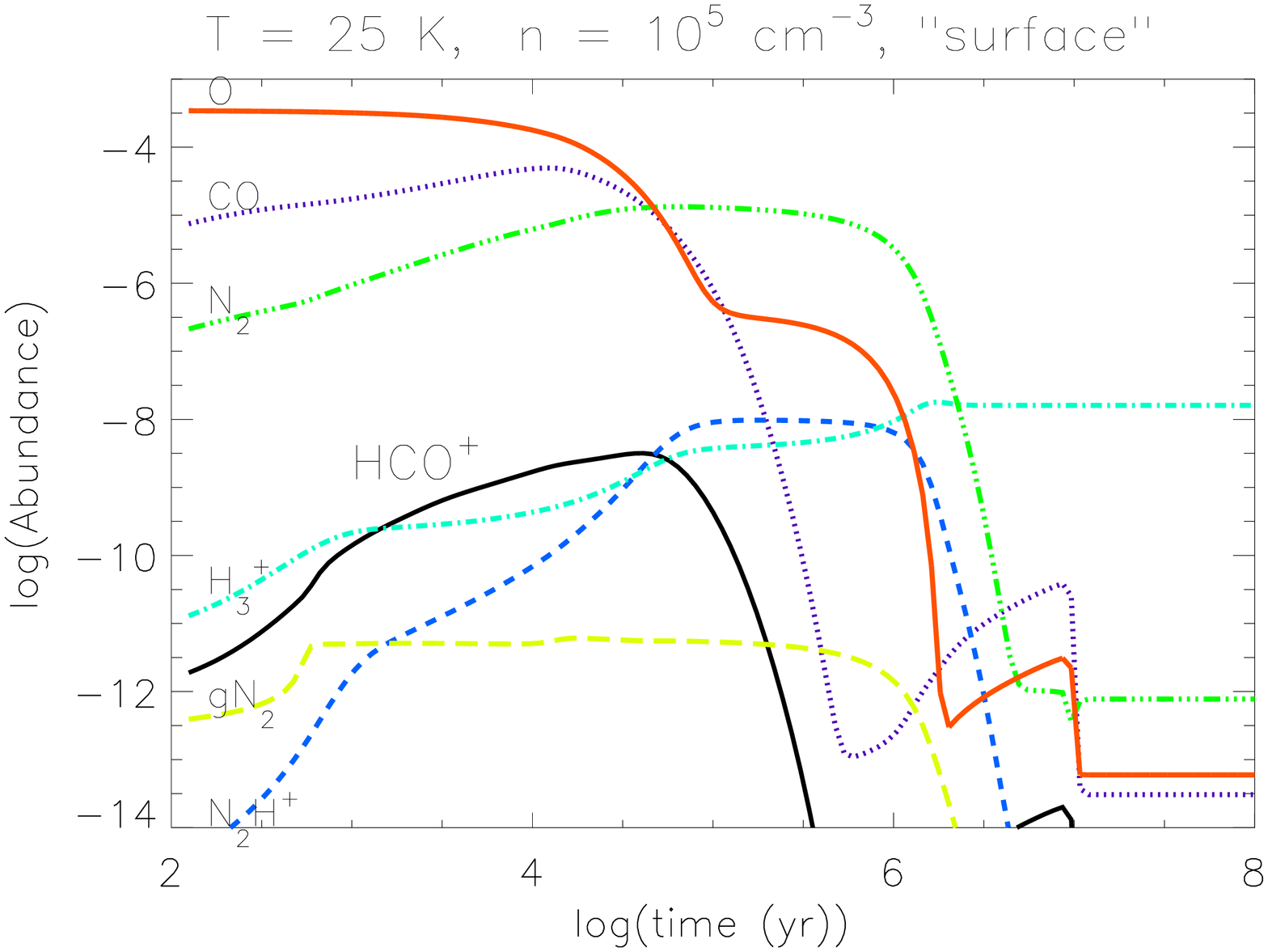}	
	
	\medskip
	\medskip
	\medskip
	\includegraphics[width=6cm,angle=90]{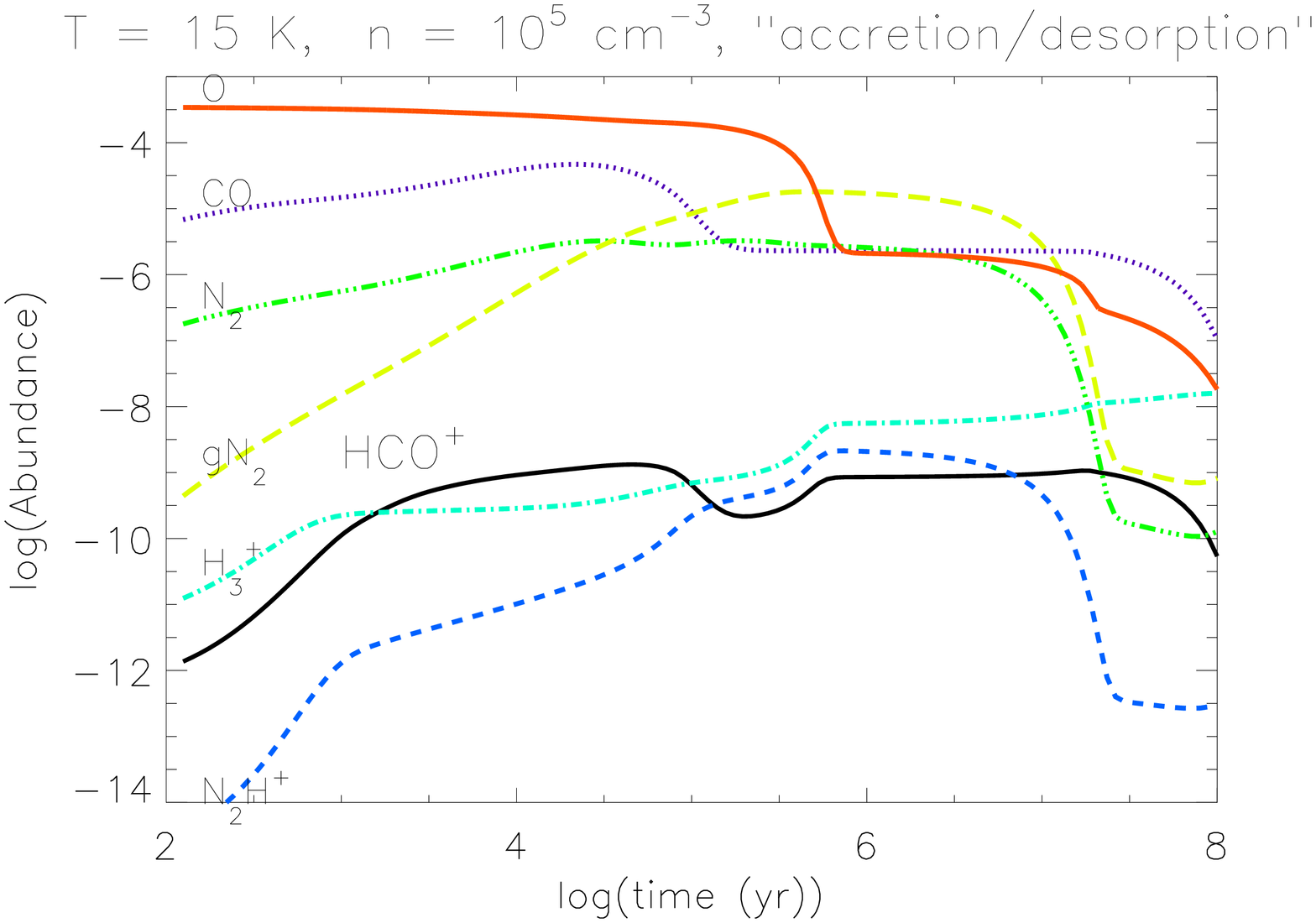}
	\includegraphics[width=6cm,angle=90]{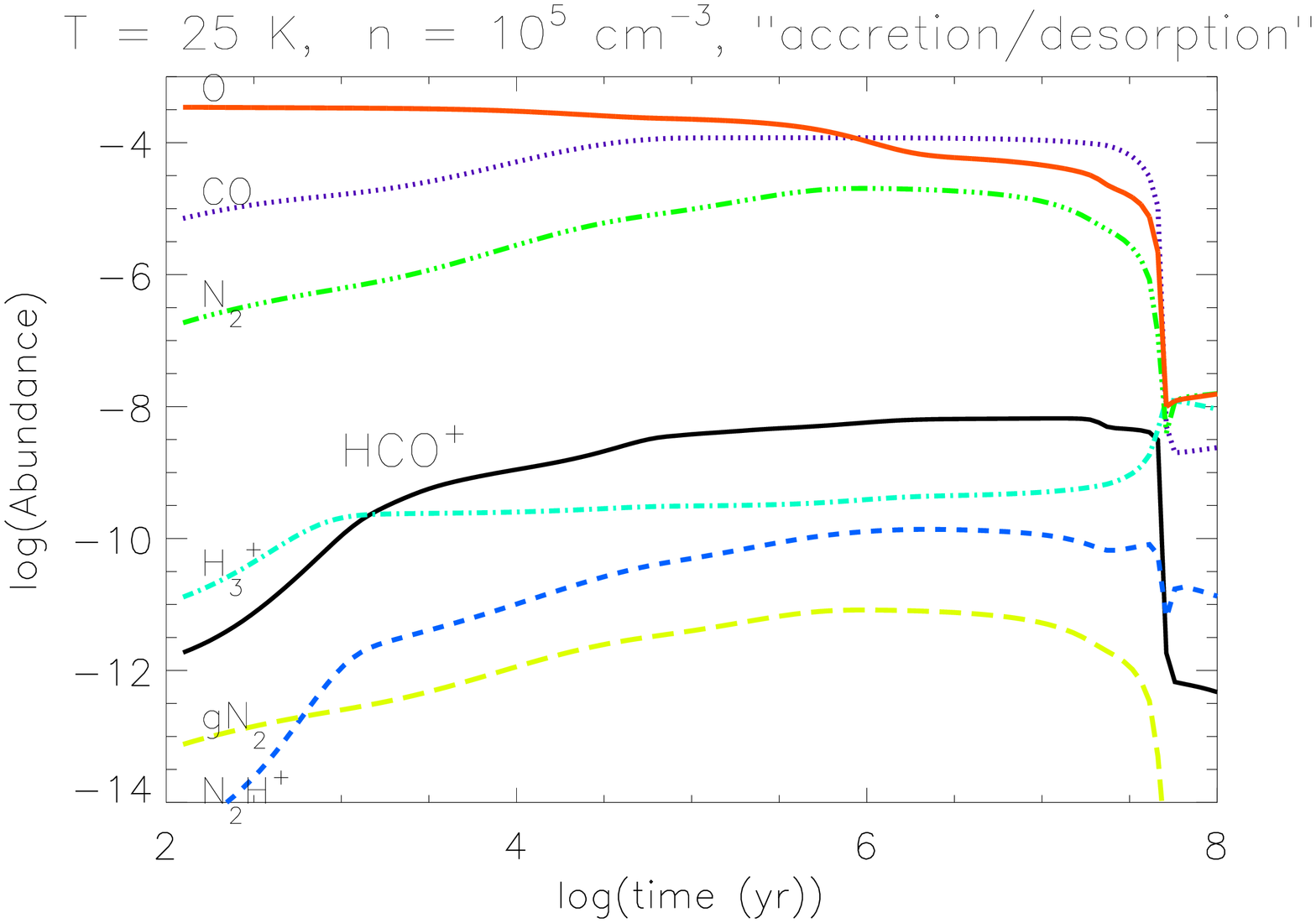}	
	\caption{Profiles of assorted species relative to H$_2$ as a function of time.  The symbol ``g" stands for a granular species.
		Upper left panel: $T$ = 15 K, $n$ =10$^5$ cm$^{-3}$, surface network.
		Upper right panel: $T$ = 25 K, $n$ =10$^5$ cm$^{-3}$, surface network.
		Lower left panel: $T$ = 15 K, $n$ =10$^5$ cm$^{-3}$, accretion/desorption network. 		
		Lower right panel: $T$ = 25 K, $n$ =10$^5$ cm$^{-3}$, accretion/desorption network.
		(A color version of this figure is available in the online journal.)}
\label{figure:suface_one}
\end{figure}

\end{document}